\def\beq{\begin{equation}}

\def\n{\nonumber}

\def\bbbz{{\bf Z}}

\def\bbb1{{\rm 1\!1}}

\newcommand{\refs}[1]{(\ref{#1})}

\documentclass[12pt]{JHEP}

\usepackage{epsfig}

\advance\hoffset by -.35in
\def\pa{\partial}

\def\ha{{1\over 2}}
\def\>{\rangle}
\def\<{\langle}
\def\mtx#1{\quad\hbox{{#1}}\quad}

\def\D{\tilde D}
\def\A{{\cal A}}
\def\C{{\cal C}}

\def\be{\beta}

\def\al{\alpha}

\def\Tr{\hbox{Tr}}

\def\be{\beta}

\def\al{\alpha}

\def\nn{\nonumber}

\def\gy{g_{\hbox{\tiny YM}}}
\preprint{KCL-TH-00-57}

\title{\Large\bf String Loop Corrections to Stable Non-BPS Branes}

\author{N.D. Lambert${}^a$ and I. Sachs${}^b$\\

\vskip 24pt

$^{a}$Dept. of Mathematics\\
King's College\\
The Strand\\
London, WC2R 2LS\\
England\\

\vskip 24pt

$^{b}$Theoretische Physik\\
Ludwig-Maximilians Universit\"at\\
Theresienstrasse 37 \\80333 Munich\\ Germany }

\abstract{
We calculate the string loop corrections to the tachyon potential for 
stable non-BPS $D p$-branes on the orbifold 
${\bf T}^4/{\bf Z}_2$. 
We find a non-trivial phase structure and we show
that, after tachyon condensation, the non-BPS $D p$-branes are
attracted to each other  for $p=0,1,2$. 
We then identify the 
corresponding closed string boundary states together with the massless 
long range fields they excite. For $p=3,4$ the string loop correction 
diverge. We identify the massless closed string fields responsible for 
these 
divergencies and regularise the partition function using a 
Fischler-Susskind mechanism.}

\begin{document}

\section{Introduction}

Non-BPS states provide an important testing ground for various
dynamical aspects of non-perturbative string theory which are
not protected by supersymmetry. A peculiar feature of non-BPS branes 
is that they typically include tachyons among their worldvolume fields. 
Much has been learned about the fate of these tachyons, both with 
conformal field theory methods, where they correspond to marginal 
deformations of the open string sigma model 
\cite{Sen,Sen9904207,9805019,Sen9805170,Sen9808141,Sen9812031,JJ,
Sen0003124,MJ,Sen0009090} and also within open string field 
theory where the tachyons do not correspond to marginal deformations 
\cite{Samuel,Sen9912249,Wati0001201,HK,Moeller0002237,Berkovits0001084,
Berkovits0002211,dMJMT,Moeller0005036,Witten,Sen9911116}. So far these
results have mainly been restricted 
to tree level approximation\footnote{however see \cite{JJ,PITB} for 
some aspects 
of tachyon condensation in the boundary state formalism}. 
On the other hand it is important to understand string loop corrections to 
these classical results, especially in view of a supergravity description of 
non-BPS branes \cite{EP,L,Oz,DiVecchia}. Of course, this aspect is also 
relevant to possible generalisations of the AdS/CFT correspondence 
\cite{Malda} to non-supersymmetric models. Indeed, the effective 
field theories describing the low energy excitations of non-BPS branes 
are non-supersymmetric at all scales 
\cite{Tseytlin,LS1,LS2}, not just in the infrared.  
 
In this paper we address the issue of open string-loop corrections to the 
tachyon potentials for non-BPS states. Of course, in order to have 
convergent loop integrals we should expand around a background which is 
a minimum of the tree level potential. Hence we take as our starting point 
stable non-BPS branes in type II theory (see \cite{Sen9904207} and references 
therein). The results we obtain in this way nevertheless enable us to draw 
some conclusions about non-critical non-BPS branes. For example, it 
has been shown in  \cite{GS} that, in the absence of tachyon
condensates and away from the critical radius, 
stable non-BPS branes repel. Therefore
it is not clear that a supergravity solution can be found. 
Here we will see that, at the critical radius, stable non-BPS D$p$-branes
with $p=0,1,2$ in fact choose a vacuum where their worldvolume scalars
arising from the tachyon momentum modes condense. Furthermore in this
vacuum the force between two stable and  
parallel non-BPS branes is attractive even 
away from the critical radius. 
Thus one  may expect to find supergravity solutions representing large
numbers of these non-BPS branes at weak string coupling.

The rest of this paper is organised as follows. In section two we prepare 
the ground by reviewing the definitions of properties of non-BPS
branes 
relevant to our calculations. In section three we calculate,
using a field redefinition introduced in section two, the expression
for the annulus partition function of open string theory. In section 
four we evaluate the effective potential numerically, after regularising 
the partition function. We also discuss the force 
between two parallel non-BPS branes at the minima of the effective
potential for various orbifold radii. In section five we construct
the closed string boundary states of the non-BPS branes obtained 
at the minimum of
the effective potential. This provides a check on our work and 
furthermore allows us to identify the massless 
closed string counterterms needed to
regularise the effective potential. Also, we extend the 
discussion of the existence of bound states for non-critical orbifolds. 
Finally, in section six we summarise 
and discuss the various results and discuss some open problems.

\section{Review of Stable Non-BPS Branes Wrapped on 
$T^{4}/{\cal I}_4(-1)^{F_L}$.}

Let us first review some features of non-BPS branes that we will need.
A non-BPS D$p$-brane of type IIA/B string theory can be constructed
by taking a $Dp$-$\bar D p$-brane pair in type IIB/A
string theory (see \cite{Sen9904207,Gaberdiel} for a review). 
There are then four Chan-Paton (CP) factors
corresponding to the locations of the two ends points of the open strings.
For the strings with both ends on the same brane the GSO
projection
removes the NS ground state so that the lightest modes form a 
maximally supersymmetric Yang-Mills $U(1)\times U(1)$ 
multiplet $(A_\mu,\phi^I,\lambda)$ in $p+1$ dimensions. 
However, for strings that stretch between different branes, the NS ground 
state  survives the GSO projection. Therefore the
lightest modes consist of a complex tachyon $T$  and a thirty-two component 
massless Fermion $\psi$. The next step is to mod-out the theory by
$(-1)^{F_L}$, where $F_L$ is the left-moving spacetime Fermion number.
This will take us from type IIB/A theory to type IIA/B theory in the
bulk. For the open strings this identifies the $Dp$-brane with the
$\bar D p$-brane and projects out all but the $I$ and $\sigma_1$ CP factors. 
Thus the gauge group in $I$-sector is projected
down to $U(1)$ representing a single object. In the
$\sigma_1$-sector the tachyon and Fermions become real. This results in 
a non-BPS D$p$-brane ($\D p$-brane) with $p$ odd/even 
in type IIA/B string theory. However, the presence of the tachyon
implies
that a $\D p$-brane is unstable and will decay into the vacuum.

We may construct stable $\D p$-branes by  wrapping a $\D(p+4)$-brane over
the orbifold ${\bf T}^4/g$ where 
\begin{equation}
g = {\cal I}_4 (-1)^{F_L} \ ,
\end{equation}
and ${\cal I}_4:x^i \rightarrow -x^i$, $i=6,7,8,9$. It was shown by
Sen \cite{Sen} that the NS-$\sigma_1$ ground state is odd
under $g= {\cal I}_4(-1)^{F_L}$. Therefore the tachyon zero-mode is 
projected out 
whereas the
modes with odd winding around the orbifold are not. In this way we
see that the lightest states in the NS-$\sigma_1$ sector are
the four scalar modes $\chi^i$ coming from the Fourier expansion of 
$T$
\begin{equation}
\frac{\chi^i(x^\mu)}{2i}(e^{iX^i/R_i} - e^{-iX^i/R_i})\ ,
\end{equation}
where $x^\mu$ are the non-compact coordinates along the brane.
These have a mass-squared $m_i^2 = -\frac{1}{2}+1/{R_i^2}$ which
is positive if $R_i \le \sqrt{2}$. 
Thus we obtain a stable  $\D p$-brane in the
non-compact space. Here, $p$ labels the number of non-compact
extended spatial dimensions of the brane\footnote{Often a stable 
non-BPS brane with $p$ non-compact spacial dimensions is 
defined by  taking the T-dual along the wrapped orbifold 
directions \cite{GS}. Then 
the massless scalars $\chi_i$ arise from winding modes in the orbifold 
directions.}. Note that, although we
continue to use the term ``tachyon'' to  describe various scalars
$\chi^i$, 
after the orbifold projection at the critical radius these modes are
in fact massless. Thus the lightest states in the $\sigma_1$-sector
to survive the orbifold consist of four massless scalars $\chi^i$ and
a sixteen component massless  Fermion $\psi_-$. In the $I$-sector the
effect of the orbifold is simply to remove the scalars $\phi^i$,
that represent the transverse motion along the orbifold direction, and
also to remove half of the Fermions $\lambda_+$.  In total we now
find the same field content as a maximally supersymmetric Yang-Mills
multiplet in $p+1$ dimensions: a vector $A_\mu$, $\mu=0,1,2,...,p$, 
$9-p$ scalars $\phi^I,\chi^i$, $I=p+1,...,5$, $i=6,7,8,9$, 
and two sixteen component Fermions $\lambda_+,\psi_-$. Here $\pm$
labels the six-dimensional chirality of the Fermions.

The low energy effective field theory for these non-BPS branes 
has been derived in \cite{LS1,LS2}. At the critical radius, the lowest
order term in an $\alpha'$ expansion is
\begin{eqnarray}
{\cal L}&=&\Tr\left[\frac{1}{4}F_{\mu\nu}F^{\mu\nu}
+\ha(D_{\mu}\phi_{I})(D^{\mu}\phi^{I})+\ha(D_{\mu}\chi_{i})
(D^{\mu}\chi^{i})
-{i\gy}\bar\lambda_+\Gamma^\mu D_\mu\lambda_+ \right. \nonumber\\
&&\left. \qquad
-{i\gy}\bar\psi_-\Gamma^\mu D_\mu\psi_- 
+\gy\bar\lambda_+\Gamma^I[\phi^I,\lambda_+]
+\gy\bar\psi_-\Gamma^I[\phi^I,\psi_-]- V\right] \nonumber\\
\label{lagrangian}
\end{eqnarray}
where
\begin{equation}
V=\frac{\gy^2}{4} \sum_{I,J} ([\phi^I,\phi^J])^2
+\frac{\gy^2}{2}\sum_{I,j} ([\phi^I,\chi^j])^2
- \frac{\gy^2}{4}\sum_{i\ne j}(\{\chi^i,\chi^j\})^2\ ,
\label{potential}
\end{equation}
and $D_\mu = \partial_\mu + i\gy[A_\mu,\ ]$.
The moduli space of vacua is parameterised by constant
scalar  vevs which satisfy 
\begin{equation}
[\phi^I,\phi^J]=[\phi^I,\chi^j]=\{\chi^i,\chi^j\}=0\ ,
\label{flatdirections}
\end{equation}
for $i\ne j$. A peculiar feature of this model is that the potential 
\refs{potential} has flat directions where two or more 
anti-commuting tachyon vevs are turned on. 
Furthermore it was shown in \cite{LS2} that these flat
directions, which correspond to marginal deformations, persist
to all orders in $\alpha'$ at tree level in open string theory. 
The purpose of this paper is to determine the fate of these flat 
directions 
after the inclusion of one loop open string  corrections.  

If we move away from the critical radius then then a mass term
for the $\chi^i$'s appear in the Lagrangian \refs{lagrangian}
\cite{LS1} and the flat directions are removed. In much 
of this paper we will be most interested in the partition function at
the critical radius. 
At this point it is possible to perform a field redefinition
that maps the tachyon vertex operator into a Wilson line 
\cite{Sen9808141,Sen9904207,Sen0003124}. 
Here we will exploit this redefinition so let us also review it.

The objects of interest are the vertex operators for the momentum
modes of $T$ in the 
$(0)$ and $(-1)$-pictures. 
We use  the conventions of \cite{LS1} and set  $\al'= g_o = 1$. 
At the critical radius
\begin{equation}
       V_T^{(-1)i} = -\frac{i}{\sqrt{2}}e^{-\Phi}
\left(e^{iX^i/\sqrt{2}}-e^{-iX^i/\sqrt{2}}\right) \otimes \sigma_1\ ,
\end{equation}
in  the $(-1)$-picture, where $\Phi$ is the Bosonised superconformal
ghost,  and 
\begin{equation}
V_{T}^{(0)i}=\frac{1}{2}\psi^{i}\left(e^{i\frac{1}{\sqrt{2}}
        X^{i}}+e^{-i\frac{1}{\sqrt{2}}X^{i}}\right)\otimes \sigma_1\ ,
\label{v1}
\end{equation}
in the $(0)$-picture. 
We  first express $V_{T}^{(0)i}$ in terms of the closed string fields 
$X_{R/L}^{i},\psi_{R/L}^{i}$ with Neumann boundary condition
\begin{eqnarray}
        X^{i}_{L}=X^{i}_{R}=\ha X^{i}\ ,\n\\
        \psi^{i}_{L}=\psi^{i}_{R}=\psi^{i}\ ,      
        \label{v2}
\end{eqnarray}
at both ends in the NS-sector. In the R-sector \refs{v2} applies at
one end and 
\begin{equation}
        \psi^{i}_{R}=-\psi^{i}_{L}\ ,
        \label{v2.1}
\end{equation}
in the other. The vertex operator \refs{v1} can be written 
as
\begin{equation}
        \frac{1}{2}\psi^{i}_{R}
        \left(e^{i\sqrt{2}X_{R}^{i}}+e^{-i\sqrt{2}X_{R}^{i}}
        \right)\  .
        \label{v2.2}
\end{equation}
Next we may  Fermionise $e^{i\sqrt{2}X_{R}^{i}}$ as 
\begin{equation}
e^{i\sqrt{2}X_{R}^{i}}=\frac{1}{\sqrt{2}}(\xi^{i}_{R}+i\eta^{i}_{R})
        \otimes\Gamma^{i}\mtx{and}
e^{i\sqrt{2}X_{L}^{i}}=\frac{1}{\sqrt{2}}(\xi^{i}_{L}+i\eta^{i}_{L})
        \otimes\Gamma^{i}\ ,
        \label{v3}
\end{equation}
where the cocycles $\Gamma^{i}$ are introduced to restore the 
correct commutation relations with the worldsheet fermions 
\cite{Sen0003124,Banks}. This can be achieved by taking for 
$\Gamma^{i}$ the generators of the $Spin(4)$-Clifford algebra and 
attaching  $\Gamma_{5}=\Gamma^{6789}$ to the worldsheet Fermions. 
We complete the  transformation by re-Bosonising as 
\begin{equation}
        \frac{1}{\sqrt{2}}\left(\xi^{i}_{R/L}\pm 
        i\psi^{i}_{R/L}\right)=e^{\pm i\sqrt{2}\tilde X^{i}_{R/L}}
        \otimes \tilde \Gamma^i
        \ , \label{v4}
\end{equation}
and attaching a $\tilde\Gamma_{5}$ to $\eta^{i}_{R/L}$. Here the 
$\tilde\Gamma^i$ form another representation of the $Spin(4)$-Clifford 
algebra
which commutes with the $\Gamma^i$ representation.
With these 
conventions the Fermionic and Bosonic currents are then related as 
\begin{equation}
        \eta^{i}_{R}\xi^{i}_{R}=i\sqrt{2}\pa X^{i}_{R}\mtx{and}
        \psi^{i}_{R}\xi^{i}_{R}=i\sqrt{2}\pa \tilde X^{i}_{R}\ .
                \label{v5}
\end{equation}
The boundary conditions for the different fields are determined as 
follows: from \refs{v1}, \refs{v2} and \refs{v3} we have 
\begin{equation}
        \xi^{i}_{L}=\xi^{i}_{R}=\xi^{i}\mtx{and} 
        \eta^{i}_{L}=\eta^{i}_{R}=\eta^{i}.
        \label{v6}
\end{equation}
In the NS-sector, where $\psi^{i}_{L}=\psi^{i}_{R}$ on both ends 
of the open string, \refs{v6} implies
\begin{equation}
        \tilde X^{i}_{L}=\tilde X^{i}_{R}=\ha\tilde X^{i}\ ,
        \label{v7}
\end{equation}
i.e. NN boundary conditions for $\tilde X^{i}$. In the R-sector, 
where $\psi^i_{R}=\psi^i_{L}$ at one end and 
$\psi^i_{R}=-\psi^i_{L}$ at the other, 
\refs{v6} implies in turn that $\tilde X^{i}_{L}=\tilde X^{i}_{R}$ 
at one end and
\begin{equation}
        \tilde X^{i}_{L}=-\tilde X^{i}_{R}\ ,
        \label{v8}
\end{equation}
at the other
i.e. ND boundary conditions for $\tilde X^i$. Using \refs{v5},\refs{v7} and
\refs{v8} we may now write
\begin{eqnarray}
        V^{(0)i}_{T}&=&i\pa_{||}\tilde X^{i}\otimes\sigma_{1}
        \otimes\Gamma_{5}\Gamma^{i}\ , \n\\
        V^{(0)i}_{T}&=&i\pa_{\perp}\tilde X^{i}\otimes\sigma_{1}
        \otimes\Gamma_{5}\Gamma^{i}\ ,
        \label{v9}
\end{eqnarray}
in the NS and R-sectors respectively. Finally the
vertex operators $V^{(-1)i}$ in the $(-1)$-picture become simply
\begin{equation}
V_T^{(-1)i} = e^{-\Phi}\eta^i\otimes \sigma_1 \otimes 
\tilde\Gamma_5\Gamma^i\ .
\end{equation} 
To summarise, in the new variables ($\tilde X^i,\eta^i$) the 
tachyon vev takes the form of a non-Abelian Wilson line in the 
NS-sector and of a shift in position in the R-sector respectively. In 
order to obtain the generalisation of \refs{v9} to non-Abelian 
tachyon vevs all we need to do is to tensor \refs{v9} with an 
element of the $u(N)$ Lie algebra generated by $t^a$. 
For example in the NS-sector we  have 
\begin{equation}
        V^{(0)ia}_{T}=\pa_{||}\tilde X^{i}\otimes t^{a}\otimes\sigma_{1}
        \otimes i\Gamma_{5}\Gamma^{i}\ .
        \label{v11}
\end{equation}

\section{The One Loop Open String Partition Function}

Here we calculate the one loop partition function for 
tachyon vevs in open string theory. 
For simplicity 
we only consider cases where a single tachyon, $\chi^9$, has a 
non-vanishing vev. The case where two distinct 
tachyons have non-vanishing (and therefore anti-commuting) vevs is 
technically more involved but we do not expect the results to change 
qualitatively (see also \cite{LS1}). 
Although we only need  the $x^9$ orbifold radius to be critical, 
in this section we will  assume that the other radii are also critical.  
We also consider
the case of two ${\tilde D}p$-branes, so that the gauge group is $U(2)$.  
Therefore there are  two cases to consider: either
$\chi^9=v_0 t^0$ or $\chi^9=v_3t^3$, where $t^0$ is the identity and
$t^1,t^2,t^3$ are the Pauli matrices.

The partition function on the annulus is given by
\begin{eqnarray}
Z &=& \int_0^\infty {dt\over 2t}\Tr_{NS-R}\left(e^{-2tH_o}{1+(-1)^F\over 2}
{1+g\over 2}\right)\nonumber\\
&=& \int_0^\infty {dt\over 2t}\Tr_{NS-R}\left(e^{-2tH_o}{1+(-1)^Fg\over 2}
\right)
\nonumber\\
\label{Zdef}
\end{eqnarray}
Here the second line follows by summing over the $I$ and $\sigma_1$
CP sectors and using the fact that the trace over $((-1)^F+g)$
vanishes \cite{GS}.

In light-cone gauge and using the conventions of \cite{GS}, 
the open string Hamiltonian is
\begin{equation}
H_o = \pi \sum_{\mu=0}^p p_\mu p^\mu + \pi\sum_{i=6}^9p_ip^i 
+ \pi \sum_{m=0,\atop m\ne  l.c.}^9\left(
\sum_{n>0}\alpha^m_{-n}\alpha_n^m+\sum_{r>0} r\psi_{-r}^m \psi_r^m\right) 
+\pi C_o\ ,
\end{equation}
where the third sum is only over the non-light-cone
coordinates. By integrating over the momentum in the non-compact
directions we arrive at the following expression
\begin{eqnarray}
Z &=& A_p
\int_0^\infty {dt\over t^{(p+3)/2}} {\cal A}\nonumber \\
A_p &=& {1\over 2^{(p+3)/2}}\int d^{p+1}\left({k\over 2\pi}\right) \ 
,\nonumber\\
\end{eqnarray}
where $\A$ is the oscillator sum over all the $U(2)$ CP factors
(but not the $I$ and $\sigma_1$ CP factors which were already summed in
\refs{Zdef}) and the projector $\frac{1}{2}(1+ (-1)^F g)$.

We now consider the annulus correction to the tree-level tachyon 
potential. 
The idea is  to compute the annulus partition function in terms 
of the new fields ($\tilde X^i,\eta^i$) 
where the tachyon background takes the 
form of a non-Abelian Wilson line. 
We will choose the 
variables $X^m,\; m\in\{0,\cdots,8\}$ and $\tilde X^9$ instead of 
$X^9$, that is we change variables only in the $X^9$-direction. Similarly
we only change the Fermionic mode $\psi^9$ to $\eta^9$.
We begin by listing the transformations of the various fields 
under $(-1)^F$ and $g$. We have 
\begin{eqnarray}
  \label{L00}
 \hbox{ {\underline{(-1)$^F$}}}:&&\n\\
&&X^m\to X^m;\qquad \psi^m\to-\psi^m,\qquad m=0,\cdots,8\nn\\
&&\eta^9\to \eta^9;\qquad \tilde X^9\to -\tilde X^9,\n\\
&&\\
 \hbox{ {\underline g}}:&&\n\\
&&X^m\to X^m;\qquad \psi^m\to \psi^m,\qquad m=0,\cdots,5\nn\\
&&X^j\to -X^j;\qquad \psi^j\to -\psi^j,\qquad j=6,\cdots,8\nn\\
&&\eta^9\to -\eta^9;\qquad\tilde X^9\to 
-\tilde X^9.\n\\  
\end{eqnarray}

First we note 
that the $R$-sector doesn't couple to the tachyon Wilson line. 
This is already suggested by the tree-level field theory Lagrangian 
\refs{lagrangian} since no Fermions couple to the scalars $\chi^i$. 
That this is the case in full string theory follows from the 
observation that the world-sheet fields $\tilde X^9$ satisfy ND boundary 
conditions in the R-sector. As a result, both, the momentum and the 
winding 
number vanish in that sector. Consequently  a Wilson line
representing a tachyon vev has no effect on the Ramond sector. Due
to Fermionic zero-modes there is also no contribution from the 
$(-1)^Fg$ term in the open string trace and we find  (including a factor
$4$ from the CP labels $t^a$) 
\begin{equation}
4\A_R = -{2}^{3/2}{f_2^7f_3\over f_1^7f_4}
\prod_{i=6}^8\left(\sum_{n_i} q^{n_i^2}\right)
= -8 {f_2^4 f_3^4\over f_1^4 f_4^4}\ .
\label{ZR}
\end{equation} 
Here $f_1(q),f_2(q),f_3(q),f_4(q)$ are the usual oscillator functions
(for example see \cite{GS}) and $q = e^{-\pi t}$.
In \refs{ZR} we used the fact that seven Bosons and Fermions are 
unaffected,
whereas one Boson and one Fermion have the opposite modings and
that there is no momentum around $x^9$.

On the other hand, if no tachyon vev is turned on the NS sector 
contribution reads
\begin{equation}
4\A_{NS}=2{f_3^8\over f_1^8} \prod_{i=6}^9\left( \sum_{n_i}q^{n_i^2}\right)
-2{f_3^3f_4^5\over f_1^5 (\frac{1}{\sqrt{2}}f_2)^3}
\left(\sum_{n_9} q^{n_9^2}\right)
= 8{f_3^{12}\over f_1^4 f_2^4 f_4^4}- 8{f_3^4 f_4^4\over f_1^4 f_2^4} \ .
\end{equation}
Here the first term simply comes from eight Bosons and Fermions with
the usual modings and momentum around the orbifold directions. The
second terms arises because three Bosons and five Fermions 
change sign under $(-1)^Fg$. In addition  there is only one momentum sum 
because the zero-modes in the $x^6,x^7,x^8$ directions are projected out.
Therefore, at the critical radius and in the absence of tachyon vevs, 
$4\A_{NS}+4\A_R=0$ and hence the one-loop
partition function vanishes for all $t$ \cite{GS}.

To continue we follow \cite{Sen9808141,Sen9812031} and 
define $(-1)^{\tilde F}$ and $(-1)^{\tilde n_9}$ by 
\begin{eqnarray}
  \label{LL11}
  (-1)^{\tilde F}&:& \psi^a\mapsto -\psi^a\mtx{for} a\neq 9\n\\
  &&\eta^9 \mapsto -\eta^9\\
  &&X^a\mapsto X^a\mtx{;} \tilde X^9\mapsto \tilde X^9\n\\
  (-1)^{\tilde n_9}&:& \psi,\eta\mapsto\psi,\eta\n\\
  &&e^{i\sqrt{2}(n_mX^m+\tilde n_9\tilde X^9)}\mapsto (-1)^{\tilde n_9}
  e^{i\sqrt{2}(n_mX^m+\tilde n_9\tilde X^9)}
\mtx{(no summation on $m\ne9$)}.\n
\end{eqnarray}
Consequently we have\footnote{The action of $ (-1)^{\tilde F}(-1)^{\tilde 
n_9}$ on the Fock vacuum $|0> $ is defined as $(-1)^{\tilde F}(-1)^{\tilde 
n_9}|0>=(-1)^F(-1)^{n_9}|0>=-|0>$.}
\begin{equation}
  \label{LL11.2}
  (-1)^F(-1)^{n_9}=(-1)^{\tilde F}(-1)^{\tilde n_9}.
\end{equation}
Note, however, that individually $(-1)^{\tilde F}$ and $(-1)^{\tilde n_9}$ 
act quite differently from $(-1)^{ F}$ and $(-1)^{n_9}$. In particular, 
$n_9=1,F= 0$ is equivalent to $\tilde F= 1,\tilde n_9= 0$. 
Indeed $n_9= 1$ corresponds to a Wilson line in the dual variable 
$\tilde X^9$, which has $\tilde F= 1$ and $\tilde n_9= 0$. Loosely 
speaking (i.e. at lowest level) the change of variables  $(\psi^9,X^9)
\to (\eta^9,\tilde X^9)$ acts like a $x^9$-momentum $\leftrightarrow$ 
fermion number duality 
transformation.   

In order to obtain the remaining 
non-vanishing couplings we first need to identify the various cocycles 
introduced in the field redefinition. The rules for 
attaching cocycles to the various operators are straightforward 
generalisations of those given in \cite{Sen9812031,Banks}.
A cocycle $\Gamma_5$ 
is attached to all $(-1)^F$-odd operators and a cocycle 
$\Gamma^9$ is attached to $(-1)^{n_9}$-odd states, where $n_9$ is the momentum 
along $X^9$ in units of $\sqrt{2}$. This cocycle is to restore the usual 
commutation relations after Fermionising and re-Bosonising 
in the $x^9$-direction. As usual, a state will be neutral with respect to 
$\chi_9$ if its vertex operator 
commutes with the
the Wilson line  $v_at^a\otimes \sigma_1\otimes \Gamma_5\Gamma^9$
and charged otherwise.
As we shall explain in the next two subsections, in order to determine the 
contribution to the annulus partition function 
from the charged fields we need to introduce a projection operator 
$\ha(1-(-1)^{\tilde F}(-1)^{\tilde n_9})$ into the open string trace. 
That is the operators  
\begin{equation}
  \label{Cproj}
  P^{\pm}=\frac{1}{4}(1+(-1)^Fg)(1\pm (-1)^{\tilde F}(-1)^{\tilde n_9})\ ,
\end{equation}
project onto the charged or uncharged states depending on whether
$CP\otimes cocyle$ factors of a state anti-commute or commute with
the  Wilson line $v_at^a\otimes \sigma_1\otimes \Gamma_5\Gamma^9$.

\subsection{Abelian Vacuum Expectation Values}

Let us now  consider in detail the case
$\chi_9=\frac{1}{2\sqrt{2}}v_0 t^0$, 
$\chi_j= 0,\;j\neq 9$. Since $t^0$ commutes with all the $t^a$'s 
we find the $\chi_9$-charges as
in table 1.

\vbox{
\medskip
\begin{center}
\begin{tabular}{|c|c|c|c|}
\hline\label{LAtable}
$(-1)^F$ & $(-1)^{n_9}$ & CP$\otimes$cocycle &$\chi_9$-charge \\
\hline
even & even & $t^a\otimes I\otimes I$& $0$ \\
\hline
even & odd & $t^a\otimes I\otimes \Gamma^9$& $2$ \\
\hline
odd & even & $t^a\otimes \sigma_1 \otimes \Gamma_5$& $2$ \\
\hline
odd & odd & $t^a\otimes \sigma_1 \otimes \Gamma_5\Gamma^9$& $0$ \\
\hline
\end{tabular}
\end{center}
Table 1: Summary of cocycles and the absolute value of the $\chi_9$-charges for an Abelian vev
$v_0t^0$. 
\medskip
}

Substitution of \refs{Cproj} into the NS open string trace leads to the 
expressions 
\begin{eqnarray}
\A_1(v_0,t)= &&\frac{1}{4}\frac{f_3^8}{f_1^8}\left(\prod\limits_{j=6}^8
    \sum\limits_{n_j\in\bbbz}q^{n_j^2}\right)
  \sum\limits_{\tilde n_9\in\bbbz}q^{(\tilde n_9+v_0)^2}-
  \frac{1}{4}\frac{f_4^5f_3^3}{f_1^5(\frac{1}{\sqrt{2}}f_2)^3}
  \sum\limits_{\tilde n_9\in\bbbz}q^{(\tilde n_9+v_0)^2}\\
 &&+\frac{1}{4}\frac{f_4^8}{f_1^8}\left(\prod\limits_{j=6}^8
    \sum\limits_{n_j\in\bbbz}q^{n_j^2}\right)
  \sum\limits_{\tilde n_9\in\bbbz}(-1)^{\tilde n_9}q^{(\tilde n_9+v_0)^2}
- \frac{1}{4}\frac{f_4^3f_3^5}{f_1^5(\frac{1}{\sqrt{2}}f_2)^3}
  \sum\limits_{\tilde n_9\in\bbbz}(-1)^{\tilde n_9}q^{(\tilde 
n_9+v_0)^2},\n
  \label{LA1}
\end{eqnarray}
for the charged states, and 
\begin{eqnarray}
\A_0(t)= &&\frac{1}{4}\frac{f_3^8}{f_1^8}
\left(\prod\limits_{j=6}^8
    \sum\limits_{n_j\in\bbbz}q^{n_j^2}\right)
  \sum\limits_{\tilde n_9\in\bbbz}q^{\tilde n_9^2}-
  \frac{1}{4}\frac{f_4^5f_3^3}{f_1^5(\frac{1}{\sqrt{2}}f_2)^3}
  \sum\limits_{\tilde n_9\in\bbbz}q^{\tilde n_9^2}\\
 &&-\frac{1}{4}\frac{f_4^8}{f_1^8}\left(\prod\limits_{j=6}^8
    \sum\limits_{n_j\in\bbbz}q^{n_j^2}\right)
  \sum\limits_{\tilde n_9\in\bbbz}(-1)^{\tilde n_9}q^{\tilde n_9^2}
+ \frac{1}{4}\frac{f_4^3f_3^5}{f_1^5(\frac{1}{\sqrt{2}}f_2)^3}
  \sum\limits_{\tilde n_9\in\bbbz}(-1)^{\tilde n_9}q^{\tilde n_9^2},\n
  \label{LA2}
\end{eqnarray}
for the uncharged states. The various terms in \refs{LA1} and \refs{LA2} 
correspond to NS, NS$(-1)^Fg$, NS$(-1)^{\tilde F}(-1)^{\tilde n_9}$ 
and 
NS$(-1)^Fg(-1)^{\tilde F}(-1)^{\tilde n_9}$ respectively. In the first 
term 
there is no change with respect to the ``old variables''. 
In the second term we have 
used that five Fermions and three Bosons change sign 
under $(-1)^Fg$ and furthermore that the momentum 
modes in $x^6,x^7,x^8$ are projected out by $g$ (as in \cite{GS}, 
but not for $x^9$).  The third and the fourth terms are then obtained 
by noting that $(-1)^{\tilde F}$ changes the sign of all Fermions 
without any effect  on the Bosons. 
  
Collecting the different terms we then end up with
\begin{equation}
 \A = 4\A_0(t) +4\A_1(v_0,t)+4\A_R(t) = \A_A(v_0,t)-\A_A(0,t) \ ,
 \label{LA3}
\end{equation}
where $\A_A(v_0,t) = A_1(v_0,t)$ and  we have included a factor of $4$ 
from 
the CP factors $t^a$.  
This second equality arises from the fact that 
at $v_0=0$ the oscillator sum $\A$ vanishes identically. For later use we 
also write
$\A_A$ in terms of $\theta$-functions
\begin{eqnarray}
\A_A(v_0,t)&=&e^{-\pi v^2_0t}\eta^{-12}(it)
\theta_3^3(0,it)\left[\theta_2^4(0,it)\theta_3(itv_0,it)\right.\nonumber\\
&&\left. \qquad+\theta_4^3(0,it)(\theta_4(0,it)-\theta_3(0,it))
\theta_4(itv_0,it)\right]\ .
\label{LA4}
\end{eqnarray}

As a check on our result \refs{LA4} let us find the 
lightest modes which will appear
in the field theory approximation.
In field theory, the only charged states are the three tachyons 
$\chi_j,\;j\neq 9$, that is, the states in the $\sigma^1$-sector with 
$n_9= 0, (-1)^F$-odd, but $n_j= \pm 1$ for one $j\neq 9$.  On the 
other 
hand, from the field theory calculation \cite{LS1} we expect $5$ massless, 
neutral  states; $4$ from the $I$-sector ($(-1)^F$-even) and one from the 
$\sigma^1$-sector with $n_9=1,\; n_j=0,j\neq 9$. We also expect
eight massless fermions which arise from the Ramond sector. Each of
these fields also carries 4 degrees of freedom coming from the
CP factors of $U(2)$. 
To obtain these states from the string partition function we need to
evaluate the leading order large $t$ behaviour for small $v_0$. 
From the Ramond sector we find 
\begin{equation}
4\A_R = -32+{\cal O}(q) \ ,
\end{equation}
whereas the NS sector gives
\begin{equation}
4\A_0 + 4\A_1 = 20+12q^{v_0^2}+{\cal O}(q) \ ,
\end{equation} 
i.e. $32$ massless Fermions, $20$ massless Bosons and $12$ massive Bosons.
Thus we obtain the correct spectrum of light states. 

\subsection{Non-Abelian Vacuum Expectation Values}

Let us now  consider in detail the case
$\chi_9=\frac{1}{2\sqrt{2}}v_3 t^3$, 
$\chi_j= 0,\;j\neq 9$. Since $t^3$ does not commute with the
other $t^a$'s we find the $\chi^9$ charges given 
in table \refs{LNAtable}.

\vbox{
\medskip
\begin{center}
\begin{tabular}{|c|c|c|c|}
\hline\label{LNAtable}
$(-1)^F$ & $(-1)^{n_9}$ & CP$\otimes$cocycle&$\chi_9$-charge \\
\hline
even & even & $t^{0,3}\otimes I\otimes I$& $0$ \\
\hline
even & odd & $t^{0,3}\otimes I\otimes \Gamma^9$& $2$ \\
\hline
odd & even & $t^{0,3}\otimes \sigma \otimes \Gamma_5$& $2$ \\
\hline
odd & odd & $t^{0,3}\otimes \sigma \otimes \Gamma_5\Gamma^9$& $0$ \\
\hline
even & even & $t^{1,2}\otimes I\otimes I$& $2$ \\
\hline
even & odd & $t^{1,2}\otimes I\otimes \Gamma^9$& $0$ \\
\hline
odd & even & $t^{1,2}\otimes \sigma \otimes \Gamma_5$& $0$ \\
\hline
odd & odd & $t^{1,2}\otimes \sigma \otimes \Gamma_5\Gamma^9$& $2$ \\
\hline
\end{tabular}
\end{center}
Table 2: Summary of cocycles and the absolute value of the $\chi_9$-charges for a non-Abelian
vev $v_3t^3$. 
\medskip
}

We see that the $t^0$ and $t^3$ sectors contribute exactly the same
$\A_0+\A_1$ terms to the oscillator sum that appeared in the case of
an Abelian vev, whereas the $t^1$ and $t^2$
sector states contribute similar terms to \refs{LA1} and \refs{LA2} but
with opposite projectors
\begin{eqnarray}
  \label{LNA1}
\A'_1(v_3,t)= &&\frac{1}{4}\frac{f_3^8}{f_1^8}\left(\prod\limits_{j=6}^8
    \sum\limits_{n_j\in\bbbz}q^{n_j^2}\right)
  \sum\limits_{\tilde n_9\in\bbbz}q^{(\tilde n_9+v_3)^2}-
  \frac{1}{4}\frac{f_4^5f_3^3}{f_1^5(\frac{1}{\sqrt{2}}f_2)^3}
  \sum\limits_{\tilde n_9\in\bbbz}q^{(\tilde n_9+v_3)^2}\\
 &&-\frac{1}{4}\frac{f_4^8}{f_1^8}\left(\prod\limits_{j=6}^8
    \sum\limits_{n_j\in\bbbz}q^{n_j^2}\right)
  \sum\limits_{\tilde n_9\in\bbbz}(-1)^{\tilde n_9}q^{(\tilde n_9+v_3)^2}
+ \frac{1}{4}\frac{f_4^3f_3^5}{f_1^5(\frac{1}{\sqrt{2}}f_2)^3}
  \sum\limits_{\tilde n_9\in\bbbz}(-1)^{\tilde n_9}q^{(\tilde 
n_9+v_3)^2},\n
\end{eqnarray}
and 
\begin{eqnarray}
  \label{LNA2}
\A'_0(t)= &&\frac{1}{4}\frac{f_3^8}{f_1^8}\left(\prod\limits_{j=6}^8
    \sum\limits_{n_j\in\bbbz}q^{n_j^2}\right)
  \sum\limits_{\tilde n_9\in\bbbz}q^{\tilde n_9^2}-
  \frac{1}{4}\frac{f_4^5f_3^3}{f_1^5(\frac{1}{\sqrt{2}}f_2)^3}
  \sum\limits_{\tilde n_9\in\bbbz}q^{\tilde n_9^2}\\
 &&+\frac{1}{4}\frac{f_4^8}{f_1^8}\left(\prod\limits_{j=6}^8
    \sum\limits_{n_j\in\bbbz}q^{n_j^2}\right)
  \sum\limits_{\tilde n_9\in\bbbz}(-1)^{\tilde n_9}q^{\tilde n_9^2}
- \frac{1}{4}\frac{f_4^3f_3^5}{f_1^5(\frac{1}{\sqrt{2}}f_2)^3}
  \sum\limits_{\tilde n_9\in\bbbz}(-1)^{\tilde n_9}q^{\tilde n_9^2}.\n
\end{eqnarray}
The total contribution to the partition function from the
oscillators is thus given by 
\begin{eqnarray}
\A(v_3,t) &=& 2\A_0(t) + 2\A_0'(t) + 2\A_1(v_3,t) + 2\A_1'(v_3,t)+ 
4\A_R(t) 
\nonumber\\
&=&
\A_{NA}(v_3,t)-\A_{NA}(0,t)
\ ,
\end{eqnarray}
where $\A_{NA}(v_3,t) = 2\A_1(v_3,t)+2\A_1'(v_3,t)$. Again the
last equality arises as a consequence of the fact that $\A(0,t)=0$.
In terms of $\theta$-functions $\A_{NA}(v_3,t)$ takes the form  
\begin{equation}
\A_{NA}(v_3,t) = e^{-\pi v^2_3t}\eta^{-12}(it)
\theta_3^3(0,it)\theta_2^4(0,it)\theta_3(itv_3,it)\ .
\end{equation}
Again we can check our result by finding the lightest modes which will appear
in the field theory approximation. When counting the massless degrees 
of freedom in the presence of a 
non-Abelian vev we need to take the Higgs mechanism into account. On the 
other hand we expect thirty-two massless fermions from the 
Ramond sector. 
To obtain these states from the string partition function we need to
evaluate the leading order large $t$ behaviour for small $v_3$. 
From the Ramond
sector we find
\begin{equation}
4\A_R = -32+{\cal O}(q) \ ,
\end{equation}
whereas the NS sector gives
\begin{equation}
\A_0 + \A_1 + \A_0' +\A_1'= 16+16q^{v_0^2}+{\cal O}(q) \ ,
\end{equation}
i.e. $32$ massless Fermions, $16$ massless Bosons and $16$ massive Bosons. 
This agrees with the counting in the field theory 
Lagrangian \refs{lagrangian}.

\section{The Effective Potential}

With the preparations of section three we are now ready to evaluate the 
effective 
tachyon potential, 
\begin{equation}
V = -Z\ .
\end{equation}
Before proceeding we note that the integrand $\A$ behaves at small
$t$ as
\begin{eqnarray}
\A &=& 4t^2 ({\rm cos}(\pi v_0) - 1)({\rm cos}(\pi v_0)-3) 
= t^2\C_{A}\ ,\nonumber\\
\A &=&4t^2 ({\rm cos}(\pi v_3) - 1)({\rm cos}(\pi v_3)+1) 
= t^2\C_{NA}\ ,\nonumber\\
\label{smallt}
\end{eqnarray} 
for an Abelian and non-Abelian vev respectively.
Therefore if $p\ge 3$ the integral will diverge. 
Since there are
no open string tachyons (below the critical radius) the integral 
always converges for large $t$ and $p\ge0$. Thus we need to
regulate the $t$-integral. We do this by introducing a cut-off
$\Lambda$ and a corresponding counter term ${\cal C}$ 
(defined in \refs{smallt}) 
\begin{eqnarray}\label{ct1}
V_\Lambda &=&  -A_3\left(
\int_\Lambda^\infty {dt\over t^{3}} {\cal A} 
+{\rm ln}\Lambda {\cal C}\right)\ ,\nonumber \\
V_\Lambda &=&  -A_4\left(
\int_\Lambda^\infty {dt\over t^{7/2}} {\cal A} 
-2\Lambda^{-1/2} {\cal C}\right)\ ,\nonumber \\
\end{eqnarray}
for $p=3$ and $p=4$ respectively. This is a particular application of the 
Fischler-Susskind mechanism \cite{FS}. While it is consistent to set the 
massless closed string fields to zero at tree-level in open string 
theory 
this no longer the case at string-loop level where a $\D p$-brane acts as a 
source for these fields. Ignoring this effect will generically lead to 
small $t$ divergences in the open string partition function. These 
divergences can then be compensated by perturbing the sigma model  
by  
the vertex operators corresponding to closed string fields. In fact
with this in mind, 
since a $\D p$-brane is a source for closed string fields for any $p$, 
we should also add these counter terms to the
finite cases $p=0,1,2$ 
\begin{equation}\label{ct2}
V_\Lambda = -A_p\left(
\int_\Lambda^\infty {dt\over t^{(p+3)/2}} {\cal A} 
+{2\over 3-p}\Lambda^{(3-p)/2} {\cal C}\right)\ .
\end{equation}
We may then  take the limit
$\Lambda\rightarrow 0$ and obtain a finite effective potential 
$V_{eff}$. Of course, for $p=0,1,2$ the addition of the counter
term has no effect. We will have more to say
about the closed string interpretation and a justification for
these choices of counter terms in section five.

An analytic evaluation of the one loop effective potential does not appear 
to be  possible. To continue we therefore discuss the 
numerical evaluation of these potentials. 
In the Appendix we have plotted the 
effective potential, or more precisely $V_{eff}/A_p$, as a 
function of the scalar vevs $v_0$ and $v_3$. 
Since these graphs share similar properties for $p=0,1,2$ and
$p=3,4$ let us discuss these two cases separately.

First we consider the graphs for $p=0,1,2$. 
In these cases there is no divergence in the open string integral.
In the case of an Abelian vev we see that the absolute minimum is
at $v_0=1$. This corresponds to the $\D p$-branes, which are wrapped
over the orbifold, splitting in to $D p$-brane/${\bar D}p$-brane pairs 
sitting 
at opposite ends of the orbifold\footnote{or, when also counting 
the compact dimensions, a $\D (p+4)$ brane splits into a  
$D (p+3)$/$\bar D (p+3)$-brane pair}, with all branes at one end and
anti-branes at the other.  We note that $v_0=0$ is also
a local minimum 
(this may not be obvious from the plot for $p=2$ but on closer 
examination one can
see that $V_{eff}/A_2$ reaches a maxima of about $.1$ near $v_0=.11$) 
so the corresponding states are meta-stable.
If we turn on a non-Abelian vev then both $v_3=0,1$ are true
minima. At $v_3=1$, the $\D p$-branes again split into $D p$-brane/
${\bar D}p$-brane
pairs but this time with equal number of branes and anti-branes at
each end of the orbifold.
In this configuration the oscillator sum vanishes due to Bose-Fermi
degeneracy.

Next we consider the cases $p=3,4$. 
Here the form of the effective potential is modified by the appearance
of counter terms. We see that for an Abelian vev $v_0=0$ is now
the true minima. Although we note that in the case of a $\D 3$-brane
there is a very slight local minimum at $v_0=1$. Therefore there is a
corresponding meta-stable state. If we turn on a non-Abelian vev then
for the $\D 3$-brane $v_3=0,1$ are both minima corresponding to a
Bose-Fermi degeneracy. However for the $\D 4$-brane the minima are
at $v_3=\frac{1}{2},\frac{3}{2}$ and there is no Bose-Fermi degeneracy.
Note that the change in sign in the effective potential for a
non-Abelian vev  on a $\D 4$-brane when compared to the other $\D
p$-branes is 
due to the counter term since the integrand is positive definite.
The interpretation of the corresponding states is not clear to us 
(see also section 5) and furthermore it not clear that the 
$\D 4$-brane is a consistent background for string perturbation theory.
Therefore we will not draw any conclusions from the one loop
effective potential in this case.

We note that  there appears to be a one loop tachyon mass renormalisation. 
The mass-squared of the tachyon  mode $\chi^9$ is now of the 
form\footnote{For the sake of clarity we temporarily restore 
$\alpha'$ and $g_o$.}
\begin{equation}
m_9^2 = -{1\over2 \alpha'}+ {1\over R^2} 
+{ 4g^2_o\over \alpha'}V''_{eff}(0)\ .
\end{equation}
From this we see that the one loop
corrected critical radius, i.e. where $m_9^2=0$, is
\begin{equation}
R_c = \sqrt{2\alpha'}(1+4g_0V_{eff}''(0))\ .
\end{equation}
Of course this result is only valid provided that $V_{eff}''(0)$ is
finite at $v=0$. In fact, although $V_{eff}(v)$ is infinitely
differentiable for $v\ne 0,1$, sufficiently high derivatives diverge 
at $v=0,1$ due to infrared effects in the open string channel. This is
the same divergenence that occurs in quantum field theory. 
The second derivative exists for  $p=2,3,4$ and one can
see that the one loop corrections have increased the critical radius
(except for the case of a non-Abelian vev on a $\D 4$-brane).

In section three we assumed for simplicity that that all the orbifold radii
were critical. However, in order to turn on a tachyon
vev it is only necessary that one direction, which we took to 
be $x^9$, is critical. In general, if the other radii $R_6,R_7,R_8$
are non-critical the effect in the partition function is to
replace the momentum sum
\begin{equation}
\prod\limits_{j=6}^8
    \sum\limits_{n_j\in\bbbz}q^{n_j^2}\ ,
\end{equation}
which appears in \refs{LA1}, \refs{LA2}, \refs{LNA1} and \refs{LNA2} by
\begin{equation}
\prod\limits_{j=6}^8
    \sum\limits_{n_j\in\bbbz}q^{2n_j^2/ R_j^2}\ .
\label{offcrit}
\end{equation}
It is easy to check that that $v_0=0,1$ and $v_3=0,1$ are still
extrema of the resulting effective potential. We have not
studied the resulting numerical form for the effective potential
but we expect that all the minima and
maxima which appear in the graphs at the critical radius persist
for any (stable) values of $R_6,R_7,R_8$.
Note that if any of the $R_j > \sqrt{2}$ then tachyons will be
re-appear in the open string spectrum and the $\D p$-branes will
become unstable.

Finally let us discuss the force between two $\D p$-branes for
general, but stable, values of $R_6,R_7,R_8$. In the
vacua where the tachyons have a vanishing vev this calculation was
done in \cite{GS}. However we have seen that these need not be the
true vacua of the system. Therefore let us re-consider the force
between two $\D p$-branes at a  minimum of the effective potential.
Separating
the branes in the transverse space spanned by $x^I$ corresponds to
turning on a vev $<\phi^I> = v^I$. We can see from the disk-level
effective action that this is always  a marginal deformation
in the presence of an Abelian tachyon vev $<\chi^9>=v_0 t^0$ whereas
for a non-Abelian vev $<\chi^9>=v_3t^3$ we must have
$v^I = v^I_3 t^3$. Therefore, in either case, the vev 
$\phi^I = v^I_3t^3$ that corresponds to separating the
branes is a marginal deformation. 

A separation of the branes can be mapped
to a Wilson line  $\phi^I_3t^3\otimes I \otimes I$. In the effective
potential this
corresponds to  introducing a multiplicative factor $e^{-r^2t/2\pi}$,
where $r = \sqrt{\sum_I <\phi^I>^2}$, for states that don't commute 
with the Wilson line (i.e. the $t^1,t^2$ CP sectors) whereas the
states which do commute (i.e. the $t^0,t^3$ CP sectors) are unaffected. 
The resulting form for the effective potential is
\begin{eqnarray}
V_{eff} &=&  -A_p
\int_0^\infty {dt\over t^{(p+3)/2}}
(2\A_0(t)+2\A_1(v_0,t)+2\A_R(t))e^{-r^2t/2\pi} + V_0\ ,\nonumber\\
V_{eff} &=&  -A_p
\int_0^\infty {dt\over t^{(p+3)/2}}
(2\A'_0(t)+2\A'_1(v_3,t)+2\A_R(t))e^{-r^2t/2\pi} + V'_0\ ,\nonumber\\
\label{force}
\end{eqnarray}
in the cases of an Abelian and non-Abelian tachyon vev respectively.
Here $V_0$ and $V'_0$ are the contributions from the $t^0$ and 
$t^3$ CP-sectors  which do not depend on $r$. Note that now the
Ramond sector states do couple to the relevant Wilson line.

For $p=3,4$ the integral in \refs{force} is  divergent since, for small
$t$,
\begin{eqnarray}
2\A_0(t)+2\A_1(v_0,t)+2\A_R(t)
&=&2t^2  (\epsilon {\rm cos}^2(\pi v_0)-4{\rm cos}(\pi v_0)+7\epsilon -4)
= t^2{\cal C}'_{A}\ ,\nonumber\\
2\A'_0(t)+2\A'_1(v_3,t)+2\A_R(t) 
&=& 2t^2(\epsilon {\rm cos}^2(\pi v_3) +4{\rm cos}(\pi v_3)+7\epsilon -12)
= t^2{\cal C}'_{NA}\ ,\nonumber\\
\label{smalltF}
\end{eqnarray}
where $\epsilon = \sqrt{R^2_6R^2_7R^2_8/8}$ and $\epsilon \le 1$
for stable $\D p$-branes. However, we may directly
calculate the force between two $\D p$-branes by differentiating 
\refs{force} with respect to $r$. This produces a convergent
integral for the force $F_r = -\partial V_{eff}/\partial r$. 
Furthermore, at least for large $r$, 
this integral is dominated by the region of small $t$.
Therefore we may find a good analytic approximation to the force by
simply evaluating 
\begin{eqnarray}
F_r&=&-{r\over \pi}A_p
\int_0^\infty {dt\over t^{(p+3)/2}}
t^3{\cal C}'e^{-r^2t/2\pi}
\nonumber\\
&=& -\frac{1}{2}A_p(2\pi)^{(3-p)/2}\Gamma({5-p\over 2})
{{\cal C'}\over r^{4-p}}\ ,
\end{eqnarray}
where ${\cal C}'$ is defined in \refs{smalltF} for the Abelian and
non-Abelian cases respectively.

Thus the potential corresponding
to the force  between two non-BPS branes satisfies the Laplacian in
the transverse space, as one expects for a conservative force. 
The one loop force $F_r$ is attractive when
${\cal C}'$ is positive,
repulsive when ${\cal C}'$ is negative and vanishes when ${\cal C}'=0$.

We have seen above that the only minima of the effective potential
(with the exception of $\D 4$-branes)
occur at $v_0=0,1$ or $v_3 = 0,1$. 
At $v_0=v_3=0$, ${\cal C}'_A = {\cal C}'_{NA}= 16(\epsilon-1)$ 
and the  force is repulsive, although it vanishes if all the radii are
critical  (in agreement with the result of  \cite{GS}).
However at $v_0=1$, ${\cal C}_A' = 16\epsilon$ and 
the branes attract each other. Furthermore, for $p=0,1,2$ this is
the true vacuum of the system. Thus we find that $\D p$-branes
with $p=0,1,2$ are in fact attracted to each other when they are
at the minimum of their effective potential.
In the case of an non-Abelian tachyon vev at $v_3=1$,
which is a degenerate global minimum along with $v_3=0$,
${\cal C}'_{NA}=16(\epsilon-2)$ and the force is repulsive. Note that at
$r=0$, $\epsilon=1$ 
there is a Bose-Fermi degeneracy at $v_3=1$, leading to
$V_{eff}=0$, which is removed if the branes are separated.

\section{Closed String Amplitudes}

As a independent check of our results for the tachyon potentials found in 
the  last section we now compute the closed string amplitudes at
$v_0=1$ and $v_3=1$ using the 
boundary state formalism. In particular, in qunatum field theory, the
effective potential cannot usually be trusted to predict new
non-trivial minima. However in string theory, by constructing the
corresponding closed string boundary states, we may test the open string
predictions. These boundary states will further allow us to identify the 
massless closed string fields responsible for the small $t$ divergence in 
the open string partition function in section three.
 
\subsection{Boundary State for a $Dp$ on $T^4$ and $T^4/g$}

According to Sen \cite{Sen9904207}, a $\D p$-brane at $v_0= 1$ is 
described by a pair of $Dp$-${\bar D}p$ branes attached to 
the orbifold fixed points 
$x^9=0$ and $x^9= \sqrt{2}\pi$ respectively. In order to construct the 
corresponding boundary state we first consider a BPS $D p$-brane with $p+1$ 
infinite dimensions and three dimensions 
wrapped around $T^4$. This state is given 
by 
(A ${\bar D}p$-brane is obtained by choosing the opposite sign for
the RR contribution)
\begin{eqnarray}
  \label{DpU}
|Dp,a,b\>&=&\frac{1}{\sqrt{2}}\left(|p,a,b\>_{NSNS}+|p,a,b\>_{RR}\right)\ ,
\end{eqnarray}
where $a$ and $b$ denote the location of the brane in the infinite 
transverse dimension and along $x^9$ respectively, i.e. 
\begin{eqnarray}
  \label{bfx1}
  |p,a,b\>_{NSNS}&=&{\cal{N}}\int dk_{p+1}\cdots dk_5 
\sum\limits_{w_6,w_7,w_8,\atop n_9}e^{i\frac{n_9}{R_9}b}e^{ika}|p,k,w\>_{NSNS}\
,\n\\
  |p,a,b\>_{RR}&=&4i{\cal{N}}\int dk_{p+1}\cdots dk_5 
\sum\limits_{w_6,w_7,w_8,\atop n_9}e^{i\frac{n_9}{R_9}b}e^{ika}|p,k,w\>_{RR}\
,
\end{eqnarray}
where ${\cal{N}}$ is a normalisation constant and
\begin{eqnarray}
  \label{Up}
  |p,k,w\>_{NSNS}&=&\frac{1}{\sqrt{2}}\left(|p,k,w,+\>_{NSNS}-
|p,k,w,-\>_{NSNS}\right)\ ,\n\\
    |p,k,w\>_{RR}&=&\frac{1}{\sqrt{2}}\left(|p,k,w,+\>_{RR}
+|p,k,w,-\>_{RR}\right)\ ,
\end{eqnarray}
with\footnote{After a double Wick rotation in $x^0$ and 
$x^4$, so that the directions tangential to the brane are spacelike, 
we then work in the light cone gauge $x^4\pm x^5$ (see 
\cite{9701137}). 
For $p= 4$ we undo the Wick rotation and take $x^0\pm x^1$. 
The formulas below then have to be changed accordingly. }
\begin{eqnarray}
  \label{bf1}
  |p,k,w,\eta\>_{NSNS/RR}&=&\exp\left(\sum\limits_{n=1}^\infty\left[
-\frac{1}{n}
\sum\limits_{\mu=0,\cdots,p,\atop \;\;\;\;6,\cdots,8}\al_{-n}^\mu\tilde\al_{-n}^\mu
+\frac{1}{n}\sum\limits_{\mu=p+1,\cdots,3,9}
\al_{-n}^\mu\tilde\al_{-n}^\mu\right]\right.\\
&&\left.+i\eta\sum\limits_{r>0}\left[-
\sum\limits_{\mu=0,\cdots,p,\atop \;\;\;\;6,\cdots,8}
\psi_{-r}^\mu\tilde\psi_{-r}^\mu+
\sum\limits_{\mu=p+1,\cdots,3,9}\psi_{-r}^\mu\tilde\psi_{-r}^\mu\right]
\right)|k,\eta\>^{(0)}_{NSNS/RR} \ .\n
\end{eqnarray}
Here $|k,\eta\>^{(0)}_{NSNS/RR}$ denotes the Fock vacuum in the NSNS and 
RR 
carrying momentum and winding number  $(k_{p+1},\cdots,k_5,n_9)$ and 
$(w_6,\cdots,w_8)$ respectively. In the NS sector this uniquely 
specifies the state. In the RR sector we have 
\begin{eqnarray}
  \label{bfzero}
  \psi^\mu_-|k,-\>^{(0)}_{RR}&=&0\mtx{for} \mu=0,\cdots,p,6,\cdots,8\n\\
  \psi^9_+|k,-\>^{(0)}_{RR}&=&0\mtx{for} \mu=p+1,\cdots,3,9\\
  |k,+\>^{(0)}_{RR}&=&\psi^9_-\prod\limits_{\mu=0,\cdots,p,\atop \;\;\;\;6,\cdots,8}
\psi^\mu_+|k,-\>^{(0)}_{RR}\ .\n
\end{eqnarray}

Next we wish to consider $D p$-branes wrapped over the orbifold $T^4/g$.
We begin by displaying the action of the orbifold operation $g$, that is
\begin{eqnarray}
  \label{g1}
  k^9&\mapsto& -k^9,\quad \omega^i\mapsto -\omega^i\mtx{for}
i=6,7,8,\\
  \al_n^i&\mapsto& -\al_n^i,\quad \tilde\al_n^i\mapsto 
-\tilde\al_n^i,\quad 
\psi_n^i\mapsto -\psi_n^i,\quad \tilde\psi_n^i\mapsto 
-\tilde\psi_n^i,\mtx{for}
i=6,7,8.\n
\end{eqnarray}
On the other hand ${\cal{I}}_4$ acts on the RR-sector ground state 
as 
\begin{eqnarray}
  \label{IR1}
  |k_{p+1},\cdots,k_5,n_9,\omega_6,\omega_7,\omega_8,\eta\>&\to&\\
   \prod\limits_{i=6}^9(\sqrt{2}\psi_0^i)\prod\limits_{i=6}^9
(\sqrt{2}\tilde\psi_0^i)&&|k_{p+1},\cdots,k_5,-n_9,-\omega_6,-\omega_7,
-\omega_8,\eta\>\ .\n
\end{eqnarray}
Furthermore, $(-1)^{F_L}$ changes the sign of the RR ground state so that  
\begin{equation}
  \label{IR2}
  g|k_{p+1},\cdots,k_5,n_9,\omega_6,\omega_7,\omega_8,\eta\>=
|k_{p+1},\cdots,k_5,-n_9,
-\omega_6,-\omega_7,-\omega_8,\eta\>\ .
\end{equation}
We therefore conclude that for $b=0,\sqrt{2}\pi$ the boundary states 
\refs{DpU} are g-invariant.

Let us now turn to the twisted sector boundary states. 
The analog of \refs{bf1} in the twisted sector is given by 
\begin{eqnarray}
  \label{bf2}
|T,k,\eta\>_{NSNS/RR}&=&\exp\left(\sum\limits_{n=1}^\infty\left[-\frac{1}{n}
\sum\limits_{\mu= 
0,\cdots,p,\atop \;\;\;\;6,\cdots,8}\al_{-n}^\mu\tilde\al_{-n}^\mu+\frac{1}{n}
\sum\limits_{\mu= 
p+1,\cdots,3,9}
\al_{-n}^\mu\tilde\al_{-n}^\mu\right]\right.\\
&&\left.+i\eta\sum_{r>0}\left[-\sum\limits_{\mu= 
0,\cdots,p,\atop \;\;\;\;6,
\cdots,8}\psi_{-r}^\mu\tilde\psi_{-r}^\mu+\sum\limits_{\mu= 
p+1,\cdots,3,9}
\psi_{-r}^\mu\tilde\psi_{-r}^\mu\right]
\right)|T,k,\eta\>^{(0)}_{NSNS/RR}\ .\n
\end{eqnarray}
Here the integers $r,n$ take the values 
\begin{eqnarray}
  \label{nrt1}
  n&\in&\bbbz_+\mtx{for} \mu=0,\cdots,5\n\\
   n&\in&\bbbz_+-\ha\mtx{for} \mu=6,\cdots,9\n\\
   r&\in&\bbbz_+-\ha\mtx{for} \mu=0,\cdots,5\n\\
   r&\in&\bbbz_+\mtx{for} \mu=6,\cdots,9
\end{eqnarray}
in the NSNS-sector, and 
\begin{eqnarray}
  \label{nrt2}
  n&\in&\bbbz_+\mtx{for} \mu=0,\cdots,5\n\\
   n&\in&\bbbz_+-\ha\mtx{for} \mu=6,\cdots,9\n\\
   r&\in&\bbbz_+\mtx{for} \mu=0,\cdots,5\n\\
   r&\in&\bbbz_+-\ha\mtx{for} \mu=6,\cdots,9
\end{eqnarray}
in the RR-sector. Note that the in the twisted sector the 
Fock vacuum has no momentum, nor winding number in the orbifold
directions. 
Also, due to the Fermionic zero modes in the NSNS sector the 
the Fock vacuum is defined slightly differently. Concretely, in 
the NSNS-sector we have 
\begin{eqnarray}
  \label{TNzero}
  \psi^\mu_-|T,k,-\>^{(0)}_{NSNS}&=&0\mtx{for} \mu=6,7,8\n\\
   \psi^9_+|T,k,-\>^{(0)}_{NSNS}&=&0\mtx{for} \mu=p+1,\cdots,3,9\\
  |T,k,+\>^{(0)}_{NSNS}&=&\prod\limits_{\mu=6,7,8}
\psi^\mu_+\psi^9_-|T,k,-\>^{(0)}_{NSNS}\ ,\n
\end{eqnarray}
whereas in the RR-sector we have
 \begin{eqnarray}
  \label{TRzero}
  \psi^\mu_-|T,k,-\>^{(0)}_{RR}&=&0\mtx{for} \mu=0,\cdots,p\n\\
  |T,k,+\>^{(0)}_{RR}&=&\prod\limits_{\mu=0,\cdots,p}
\psi^\mu_+|T,k,-\>^{(0)}_{RR}\ .\n
\end{eqnarray}
In $x$-space we then have 
\begin{eqnarray}
  \label{bx1}
  |T,a,\eta\>_{NSNS}&=&2\tilde{\cal{N}}\int dk_{p+1}\cdots dk_5 
e^{ika}|T,k,\eta\>_{NSNS}\ ,\n\\
  |T,a,\eta\>_{RR}&=&4i\tilde{\cal{N}}\int dk_{p+1}\cdots dk_5 
e^{ika}|T,k,\eta\>_{RR}\ ,
\end{eqnarray}
where $\tilde{\cal{N}}$ is another normalisation constant. The GSO 
invariant 
combination is then given by 
\begin{eqnarray}  \label{Uf}  
|T,a\>_{NSNS}&=&\frac{1}{\sqrt{2}}\left(|T,a,+\>_{NSNS}+|T,a,-\>_{NSNS}
\right)\n\\   
|T,a\>_{RR}&=&\frac{1}{\sqrt{2}}\left(|T,a,+\>_{RR}+|T,a,-\>_{RR}\right).
\end{eqnarray}
For a $Dp$-brane attached to the orbifold fixed plane $x^9= 
0,\sqrt{2}\pi$ 
and wrapped around the other orbifold directions we then have $8$ twisted 
sector boundary states $|T_\al \>,\al= 1,\cdots 8$. 

\subsection{Boundary States for $Dp$$\bar D p$ pairs on $T^4/g$}

We are now ready to write down the boundary state for a $Dp\bar Dp$ state 
with the $Dp$ and the ${\bar D}p$ located at the opposite points of the 
orbifold\footnote{We temporarily set $a= 0$ and delete this label from the 
boundary state.}
\begin{eqnarray}
  \label{DbarD}
  |Dp;\bar 
Dp\>&=&\frac{1}{\sqrt{2}}\left(|p,0\>_{NSNS}+|p,0\>_{RR}\right)+
\frac{1}{\sqrt{2}}\left(|p,\sqrt{2}\pi\>_{NSNS}-|p,\sqrt{2}\pi\>_{RR}
\right)\n\\
&&+\frac{1}{4}\sum\limits_{\al=1}^{8}
\left(|T_\al \>_{NSNS}+|T_\al \>_{RR}\right)-\frac{1}{4}\sum\limits_{\al= 
1}^{8}\left(|\bar T_\al \>_{NSNS}-|\bar T_\al \>_{RR}\right)\ ,
\end{eqnarray}
where $T_\al ,\bar T_\al $ denote the twisted boundary states located for the 
$Dp$ brane at $b= 0$ and the ${\bar D}p$-brane at $b=\sqrt{2}\pi$ 
respectively. The relative signs for the various twisted state
contributions are fixed by the requirement that the $Dp$ and the 
${\bar D}p$-brane have the same twisted sector RR charge 
\cite{9805019,Gaberdiel}. 

Let us now compare the closed string tree-amplitude for the boundary state 
\refs{DbarD} with the open string loop result  \refs{LA3} for 
$v_0= 1$. At the critical orbifold radius we have 
\begin{equation}
  \label{ctree1}
\<Dp;\bar Dp|e^{-lH_c}|Dp;\bar Dp\>=4{\cal{N}}^2l^{\frac{p-5}{2}}
\prod\limits_{i= 6}^8
\left(\sum\limits_{w_i\in \bbbz}
e^{-2\pi l w_i^2}\right)\frac{f_2(\tilde q)^8}{f_1(\tilde q)^8}
\sum\limits_{n_9}(-1)^{n_9}e^{-\pi l \frac{n_9^2}{2}} \ ,
\end{equation}
where  $\tilde q= e^{-2\pi l}$ and we have used the fact that the 
twisted sector gives a vanishing contribution to the amplitude. This 
follows from the fact that 
\begin{eqnarray}
  \label{vanT}
{}_{NSNS}\<T_\al,\pm|e^{-lH_c}|T_\be,\pm\>_{NSNS}&=&-{}_{RR}\<T_\al,\pm|e^{-lH_c}|T_\be,\pm\>_{RR}\ ,
\end{eqnarray}
for $\al= \be$ and vanishes otherwise \cite{9805019}. A similar
relation holds for the twisted sector states $|{\bar T}_\alpha \>$
and any amplitude between the two different sets of states always vanishes,
i.e. the only non-vanishing amplitudes are between the same two
twisted sector states.
The closed string partition function for a boundary state $|B\>$ is given by
\begin{equation}
Z = \int dl \<B|e^{-lH_c}|B\> \ .
\end{equation}
To compare this with the open string loop result \refs{LA3} we use 
the modular transformation properties (see appendix) (with $2l= 1/t$) as 
well as the duplication formula 
 \begin{eqnarray}
  \label{double}
\theta_2(4\tau)&=&\ha (\theta_3(\tau)-\theta_4(\tau)),
\end{eqnarray}
leading to
 \begin{equation}
\<Dp;\bar Dp|e^{-lH_c}|Dp;\bar Dp\>=4{\cal{N}}^2(2)^{\frac{5-p}{2}}
t^{-\frac{p-1}{2}}\eta^{-12}(q)\theta^4_4(0,q)\theta_3^3(0,q)
(\theta_3(0,q)-\theta_4(0,q))
\ .
\end{equation}
This agrees with the open string oscillator sum  $\A_A(v_0,t)-\A_A(0,t)$
at $v_0=1$ for a suitable choice of the normalisation constant $\cal N$.
Note that to compare the open and closed string  partition functions
we must also  include the factors from the change of
the moduli space measure $d l = -dt/2t^2$. 

Next we consider a non-abelian vev $v_3= 1$. This should correspond 
to a $Dp\bar Dp$ at $x^9= 0$ and a $\bar Dp Dp$ at $x^9= \sqrt{2}\pi$. 
The corresponding boundary sate is given by linear superposition. First we 
consider the situation where the two non-BPS branes sit on top of each 
other. Then 
\begin{eqnarray}
  \label{DbarDbarDD}
  |B\>&=&\sqrt{2}|p,0\>_{NSNS}+\sqrt{2}|p,\sqrt{2}\pi\>_{NSNS}\n\\
&&+\frac{1}{2}\sum\limits_{\al= 1}^{8}|T_\al \>_{RR}+
\frac{1}{2}\sum\limits_{\al= 1}^{8}|\bar T_\al \>_{RR}\ .
\end{eqnarray}
For a critical orbifold ($R_i= \sqrt{2}$), 
the corresponding closed string tree-level amplitude is then given by 
\begin{eqnarray}
  \label{treeDDDD}
\<B|e^{-lH_c}|B\>=\frac{4}{l}(8{\cal N}^2-\tilde {\cal N}^2)
\frac{f_3^4(\tilde q)f_2^4(\tilde q)}{f_4^4(\tilde q)f_1^4(\tilde q)}\ . 
\end{eqnarray}
On the other hand we have \cite{GS}
\begin{eqnarray}
  \label{NtN}
\tilde {\cal N}^2&=&\frac{16 R_9}{R_6 R_7 R_8}{\cal N}^2\ .
\end{eqnarray}
Hence, the amplitude vanishes on a critical orbifold, as expected from 
the results in section four. 

If the two non-BPS branes are separated a distance $r$ along one of the  
non-compact directions the corresponding boundary state is 
\begin{eqnarray}
  \label{DD0DDb}
&|Dp,0,0;\bar Dp,r,0;\bar Dp,0,\sqrt{2}\pi;Dp,r,\sqrt{2}\pi\>&\ .
\end{eqnarray}
The closed string tree amplitude is found to be 
\begin{eqnarray}
  \label{treeDD0DDb}
\<B|e^{-lH_c}|B\>&=&4{\cal{N}}^2l^{\frac{p-5}{2}}
\prod\limits_{i= 6}^8
\left(\sum\limits_{w_i\in \bbbz}
e^{-2\pi l w_i^2}\right)\frac{f_2(\tilde q)^8}{f_1(\tilde q)^8}
\sum\limits_{n_9}(-1)^{n_9}e^{-\pi l \frac{n_9^2}{2}} \n\\
&&+4{\cal{N}}^2l^{\frac{p-5}{2}}e^{-\frac{r^2}{4\pi l}}
\prod\limits_{i= 6}^8
\left(\sum\limits_{w_i\in \bbbz}e^{-2\pi l w_i^2}\right)\frac{f_2(\tilde q)^8}{f_1(\tilde q)^8}
\sum\limits_{n_9}e^{-\pi l \frac{n_9^2}{2}} \n\\
&&-4\tilde N^2l^{\frac{p-5}{2}}e^{-\frac{r^2}{4\pi l}}\frac{f_3^4(\tilde 
q)f_2^4(\tilde q)}{f_4^4(\tilde q)f_1^4(\tilde q)}\ .
\end{eqnarray}
At the critical orbifold radius \refs{treeDD0DDb} can be rewritten 
using
 \begin{equation}
  \label{double2}
\theta_3(4\tau)=\ha (\theta_3(\tau)+\theta_4(\tau)),
\end{equation}
leading to 
\begin{equation}
\<B|e^{-lH_c}|B\>=4{\cal{N}}^2e^{-\frac{tr^2}{2\pi}}
2^{\frac{5-p}{2}}t^{-\frac{p-1}{2}}
\eta^{-12}(q)\theta_4^4(0,q)\theta_3^3(0,q)
\left(\theta_4^2(0,q)-\theta_3^2(0,q)\right)+ \ldots\ ,
\end{equation}
where the ellipsis denotes terms which are independent of the
separation $r$.
This agrees with the open string result \refs{force}.

We note that the boundary states \refs{DbarD} and \refs{DbarDbarDD}
can exist away from the critical radii. 
Indeed, considering the small $l$ limit of the closed string
amplitudes for arbitrary $R_6,R_7,R_8,R_9$, one see that there
are no tachyons in the open string channel provided $R_i \le \sqrt{2}$
for $i=6,7,8$, as expected from section four, and $R_9\ge \sqrt{2}$.
However if none of the radii are critical then these
boundary states are not connected to the boundary state 
at $v_a=0$ by a  marginal perturbation  (on the disk). 

We may again evaluate the force between two boundary states of the
form \refs{DbarD} and \refs{DbarDbarDD} for arbitrary but stable 
values of the radii by simply  differentiating
the partition function with respect to $r$. At non-zero $r$ the
partition function for states described by \refs{DbarDbarDD}
is given in \refs{treeDD0DDb} whereas for  states 
described by \refs{DbarD} the effect on the partition function is simply to
include a  multiplicative 
factor of $e^{-\frac{r^2}{4\pi l}}$ in \refs{ctree1}. Just as in
section four, at large $r$ we may approximate the integral by taking
only
the leading order large $l$ term in the oscillator, momentum and 
winding sums.
Note that these  are independent of the radii and hence 
the only dependence on the radii enters through 
$\cal N$ and $\tilde {\cal N}$. More explicitly we find
\begin{eqnarray}
F_r &=& -{64\Gamma(\frac{5-p}{2})\over (4\pi)^{\frac{p-3}{2}}}
{{\cal N}^2\over r^{4-p}}\ , \\
F_r &=& {64\Gamma(\frac{5-p}{2})\over (4\pi)^{\frac{p-3}{2}}}
{{\cal N}^2\left(\frac{4R_9}{R_6R_7R_8}-1\right)\over  r^{4-p}}\ , \nonumber
\end{eqnarray}
for the states  \refs{DbarD} and \refs{DbarDbarDD} respectively.
Now ${\cal N}^2 \ge 0$ and hence the force between two states described
by \refs{DbarD} is attractive for all stable values of the radii. 
On the other hand, at the critical radius 
$R_i= \sqrt{2}$, the force between
two states described by \refs{DbarDbarDD} is repulsive. 
In fact it remains repulsive away from the critical radius 
as long as $R_9/R_6R_7R_8\ge 1/4$ which always the case for stable
configurations. 
The presence of a repulsive force at the critical radius 
in spite of Bose-Fermi degeneracy is an interesting and maybe unexpected 
phenomenon. 

\subsection{Interpretation of the Counter Terms}

Using the boundary states found above we can now interpret the counter 
terms introduced in section four by identifying the massless closed string 
fields exited by these boundary
states. For definiteness we consider the case of an abelian vev $v_0$.  
Consider first the boundary state for a non-BPS $Dp$ 
brane at $v_0= 0$ 
\cite{Gaberdiel}
\begin{eqnarray}
  \label{nonBPSD}
  |\tilde Dp\>&=&\frac{1}{\sqrt{2}}|p,0\>_{NSNS}
+\frac{1}{4}\sum\limits_{\al= 1}^{8}|T_\al \>_{RR}
+\frac{1}{4}\sum\limits_{\al= 1}^{8}|\bar T_\al \>_{RR}
\end{eqnarray}
In \refs{nonBPSD} only the untwisted NSNS sector and the twisted RR sector 
boundary states survive the GSO projection. At long distance, where 
the massless closed string fields dominate the interaction the twisted RR 
repulsion works against the untwisted NSNS attraction leading to a 
no-force condition at the critical radius of the orbifold \cite{GS}. 
At $v_0= 1$ on the other hand, examination of \refs{DbarD} shows that  
the massless twisted NSNS and RR closed string fields 
excited by the  $Dp$ and ${\bar D}p$ boundary 
states \refs{DbarD} live on separate orbifold fixed planes and thus do not 
interact. Correspondingly they do not contribute to the partition function 
\refs{ctree1}. This leaves us with 
the untwisted contributions from the $Dp$ and ${\bar D}p$ boundary 
states. Naively one might think that this system should be 
tachyonic. The absence of the tachyonic mode can be understood by noting 
that the $Dp$ and the ${\bar D}p$-brane are attached to different orbifold 
fixed planes and can therefore not annihilate.  
The boundary state 
\refs{DbarD} acts as a dipole source for the untwisted RR-fields and 
therefore these fields fall off too fast to be relevant at large distance. 
The NSNS-contributions of the $Dp$ and ${\bar D}p$-brane in \refs{DbarD} 
however, are additive as always. Hence the graviton and dilaton determine 
the long distance (small $t$) behaviour of ${\cal{A}}_A$ in \refs{LA3}. 
Indeed, the asymptotic behaviour of the dilaton and the components of the 
graviton for the source 
\refs{DbarD} is expressed in terms of the Green function of the Laplacian 
in the transverse space i.e. 
\begin{equation}\label{ldmphi}
\phi, h_{mn} \propto\cases{\frac{1}{r^{3-p}}\mtx{;}p\neq 3\cr
\log{r}\mtx{;}p=3}
\end{equation}
In order to have a consistent string loop expansion we then need to include 
these massless fields into the tree-level open string amplitude. To first 
order in $g_c$ this is incorporated by perturbing the flat sigma-model by the 
vertex operator \cite{FS,AT,Callan}
\begin{equation}\label{fsv}
g_cF(\Lambda)h_{mn}\pa X^m\bar\pa X^n\ ,
\end{equation}
where $F(\Lambda)=\Lambda^{(3-p)/2}$ for $p\neq 3$ and 
$F(\Lambda)=\log(\Lambda)$ for $p= 3$. The tree-level (disk) 
tadpoles induced by the perturbation \refs{fsv} then reproduce the counter 
terms in \refs{smallt},\refs{ct1} and \refs{ct2}. 
For $p= 0,1,2$ these counter terms 
vanish as $\Lambda\to 0$ and therefore become redundant 
(see \refs{ct2}). For $p= 3,4$ on the other hand, the disk tadpoles 
diverge for $\Lambda\to 0$. However, together with the divergent loop 
amplitude \refs{ct1} the total amplitude is well defined as $\Lambda\to 0$. 

The fact that the metric and dilaton have a large $r$ 
divergence for $p= 3,4$ 
raises the question of whether these states are well defined. For $p= 3$
the logarithmic divergence of the metric is in fact an artifact of 
perturbation theory. Indeed the situation here is analogous to that
for point particles in 
$2+1$-dimensional gravity. The asymptotic six-dimensional 
metric of a $\D 3$-brane 
describes a flat spacetime with a wedge cut out 
\begin{equation}
\label{m2+1}
ds^2_{(6)}=-\eta_{\mu\nu}dx^\mu dx^\nu +r^{-8Gm}(dr^2+r^2 d\theta^2)\ .
\end{equation}
Expanding 
\refs{m2+1} in $G$ reproduces the logarithmic correction \refs{ldmphi}. 
For $p= 4$ the gravitational potential (and the dilaton) increases linearly 
with the distance. It is therefore not clear to us whether a 
$\D 4$-brane is a meaningful concept beyond tree-level. 

\subsection{Comparison with Field Theory}

In this subsection we compare the running of couplings in the 
field theory \refs{lagrangian} with the running of the dilaton 
for the corresponding boundary state. First we recall that the T-duality 
of string theory relates some coupling constants that can appear 
in the effective 
Lagrangian. In addition there is a ${\bf Z}_2^4$ symmetry $\chi^i
\leftrightarrow -\chi^i$ \cite{LS1} which is a result of
momentum conservation around $x^i$ in the full string theory. 
In this way, and allowing for field
redefinitions, all but the terms involving only $\chi^i$ scalars
can be fixed to take the form in \refs{lagrangian} \cite{LS1}.
For example T-duality and gauge invariance imply that there are
no mass terms allowed for the scalars $\phi^I$. However 
such a non-renormalisation
theorem would appear in the field theory in the guise of a hierarchy problem.
The remaining terms are less restricted by these symmetries and must
have the form
\begin{equation}
\frac{1}{2}m_i^2(\chi^i)^2+g_1\sum_{i,j}(\chi^i)^2(\chi^j)^2
+g_2\sum_{i,j}(\chi^i\chi^j)^2\ .
\end{equation}
Although string theory determines
the couplings $g_1$ and $g_2$ in terms of $\gy$ at tree-level, 
leading to the potential  \refs{potential}, 
from the field theory perspective there appears to
be no symmetry that will ensure the relations between the various 
couplings is preserved after loop corrections are included. 

In what follows we will fix the ambiguities that arise at one loop in 
quantum field theory by comparing with the string theory 
results in the previous section. After eliminating the freedom of field 
redefinitions this leaves the couplings $\gy,g_1,g_2$ and 
masses $m_i$ which may run independently. Note that quantum field theory
is not in general compatible with T-duality so that in principle
there may be more independent coupling constants.
From the closed string point of view we expect
that the couplings $\gy,g_1,g_2$ can be related to some combinations 
of the dilaton $\varphi$ and components of the metric. 
We should then be able to understand the running of these couplings 
through the coupling of the massless closed string fields to the 
$\D p$. It would be interesting to gain a more thorough understanding
of this relationship but this is beyond the scope of this paper.

Let us now then consider a $\D 3$-brane at $v_a=0$ and  T-dualise
along the orbifold, so that no dimensions of the brane are
wrapped. It is then clear (e.g. see \cite{Vecchia9797968}) 
that the dilaton does not couple and can be
taken to be constant. Therefore, one expects that near $v_a=0$, in the
field theory \refs{lagrangian} $\gy$ does not run at 
one loop\footnote{Note that if we 
did not T-dualise, although
the corresponding field theory is the same, we would obtain a
non-constant dilaton.}.
This can be verified by noting that, since the Lagrangian is
so similar to that of $N=4$ Yang-Mills, only graphs with external
$\chi^i$'s or Fermions are divergent at one loop.
At $v_1=1$ the $\D 3$-brane is also described by a field theory
approximation. However this field theory is not equivalent to 
\refs{lagrangian} with $v_1=1$. In particular, the $\D 3$-brane at 
$v_1= 1$ corresponds to wrapped
$D4$/${\bar D}4$-branes with half a unit of Wilson line on one of 
them\footnote{This is the T-dual picture of the 
$D 6$/${\bar D }6$-branes at opposite ends of the orbifold described above.}. 
In this case the dilaton
is no longer constant and we expect a running of the couplings. 
Note that in all cases the open string theory fixes the mass for the
scalars $\chi^i$.

\section{Conclusion}

In this paper we have determined the open string loop corrections to
the flat directions of the tree level potential on critical non-BPS 
$D p$-branes. The resulting
effective potentials in general have multiple minima. 
Perhaps the most
notable result is the observation that for $p=0,1,2$ the global minima
occurs at a non-zero vev for the tachyonic modes. 
Moreover in these vacua parallel $\D p$-branes are attracted to
each other, whereas they repel in the  unstable vacua.
It would be interesting to repeat the analysis of \cite{DiVecchia}
for $\D p$-branes at $v_0=1$ to see whether or not a supergravity description
is to be expected. 
We also examined the coupling of closed strings using the boundary
state formalism. This allowed us to understand in greater detail
the Fischler-Susskind mechanism that is needed to regularise the
open string effective potential. 
Although we discussed some aspects of the relation between the running
of the field theory couplings and the closed string fields, a more
thorough understanding of this relation would be very desirable in
view of a non-supersymmetric generalisation of the AdS-CFT
correspondence \cite{Malda,Tseytlin}.

Let us summarise our results on the forces between two
non-BPS branes and the  effective potential of the tachyon modes. 
We found that the force between two 
non-BPS branes obtained at
$v_a=0$ is
repulsive, except at the critical radius where the force vanishes, 
as first observed in \cite{GS} and explained by Bose-Fermi
degeneracy. 
On the other hand the non-BPS branes obtained at
$v_0=1$ are attracted to each other,
whereas the non-BPS branes obtained at $v_3=1$ repel
each other (even though there is a Bose-Fermi degeneracy), 
for all values of  the critical 
radius where the branes are stable. 
Indeed, so long as only one radius is critical, the absolute
minima for $p=0,1,2$  occurs at $v_0=1,
v_3=0$. Therefore, if we consider  for example a system of $\D 0$-branes at
$v_a=0$ then, although they are
stable, they will under go a phase transition and 
tunnel into the true minima at $v_0=1,\ v_3=0$. 
Curiously, while  the $\D 0$-branes repel each other at $v_0=0$, 
in the new vacuum at $v_0=1$ they are attracted to each other.

In this paper we have concentrated on tachyon kinks along the 
$x^9$-direction. For this it is enough to set $R_9$ to the critical value. 
However, if other  radii are critical then  we can
turn on vevs for the corresponding  $\chi^i$'s, provided that they all
simultaneously anti-commute \cite{LS1,LS2}. In this case we expect local 
minima at $\chi^I_a=0,\frac{1}{2\sqrt{2}}$. In addition one may expect new 
minima with two non-commuting tachyon vevs \cite{LS1}. 
In fact the field theory approximation suggest that these vacua are 
energetically preferred \cite{LS1}. If so there is an attractive force 
between the 
branes since, from \refs{flatdirections}, 
$\phi^I_a t^a \otimes I \otimes I$
is no longer marginal (except for the centre of mass coordinate 
$\phi^I_0$). In other words, we expect that the 
corresponding boundary state is stable (at one loop) and 
two of these states are attracted to each other. 
On the other hand the field theory approximation cannot be 
trusted in this regime and it would therefore be interesting to work out the 
open string effective potential in this case.  

In closing
we note that several other combinations of non-BPS branes and stabilising 
orbifolds can be considered \cite{GS,MOT,GSt,St,MJ}.
We expect that the non-Abelian
flat directions discussed in \cite{LS1,LS2} also exist in these
cases. It would be interesting to determine the fate of tachyonic 
flat directions and the resulting
forces in these other examples.

\section*{Acknowledgements}

We would like to thank G.M.T. Watts for his initial collaboration on
this work and for many insightful comments. 
We would also  like to thank D. Lowe, 
A. Sen, S. Theisen and A. Tseytlin for helpful 
discussions. 
I.S. would like to thank the Erwin Schr\"odinger Institute in 
Vienna and also Kings College London for hospitaliy during various stages 
of this work. N.D.L. has been  supported by a 
PPARC Advanced Fellowship and  also acknowledges
the PPARC grant PA/G/S/1998/00613.

\newpage

\section*{Appendix: Figures}

\EPSFIGURE{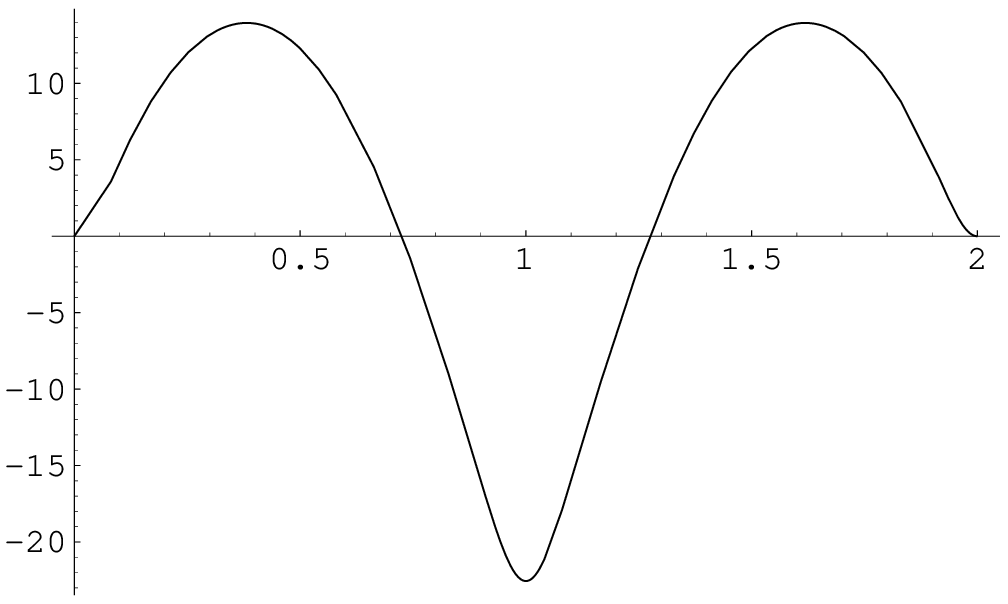}{The potential on two ${\tilde D}0$-branes
as a function of an Abelian tachyon vev.}

\EPSFIGURE{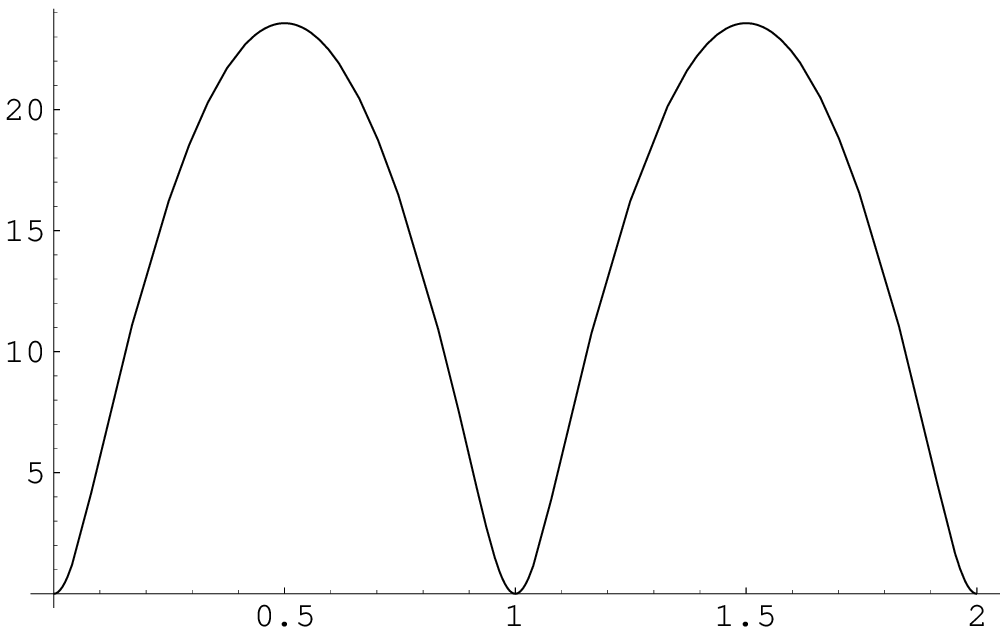}{The potential on two ${\tilde D}0$-branes
as a function of a non-Abelian tachyon vev.}

\EPSFIGURE{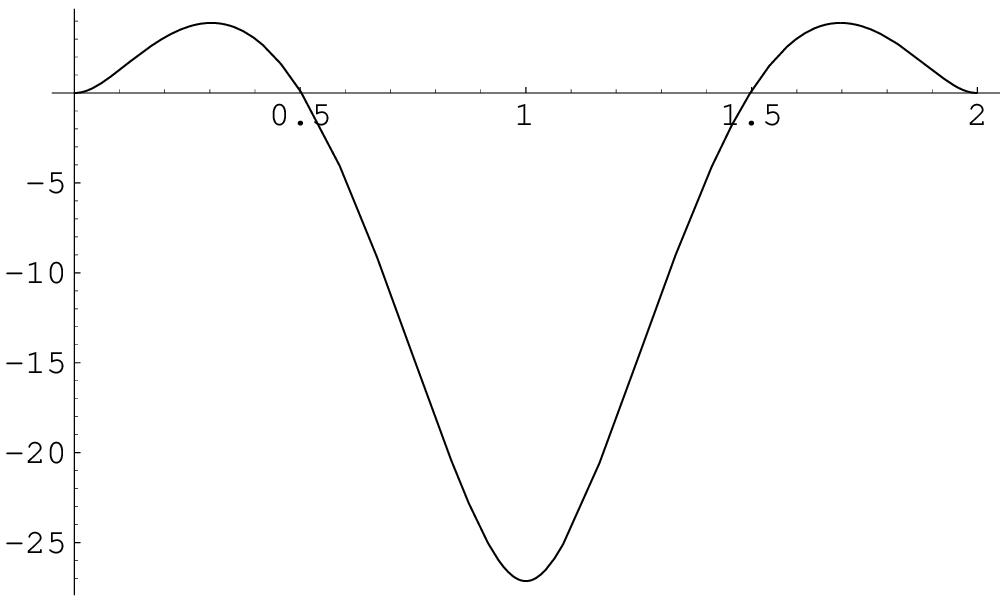}{The potential on two ${\tilde D}1$-branes
as a function of an Abelian tachyon vev.}

\EPSFIGURE{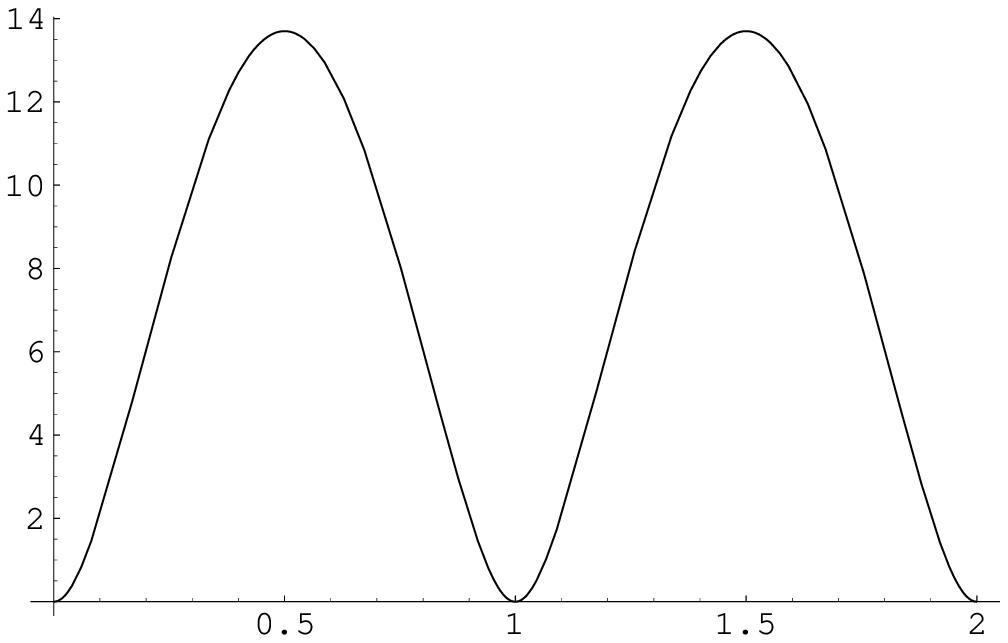}{The potential on two ${\tilde D}1$-branes
as a function of a non-Abelian tachyon vev.}

\EPSFIGURE{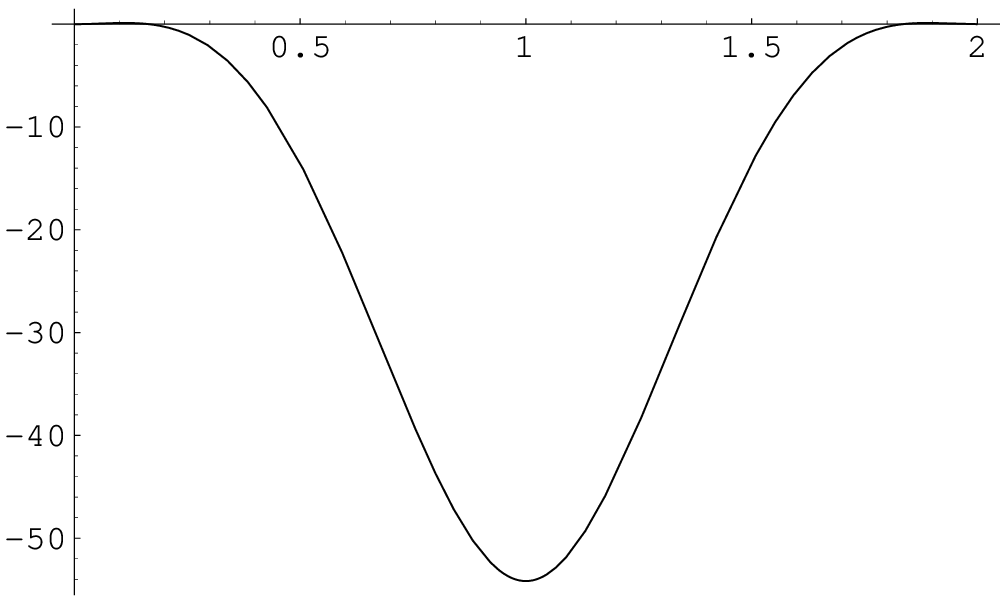}{The potential on two ${\tilde D}2$-branes
as a function of an Abelian tachyon vev.}

\EPSFIGURE{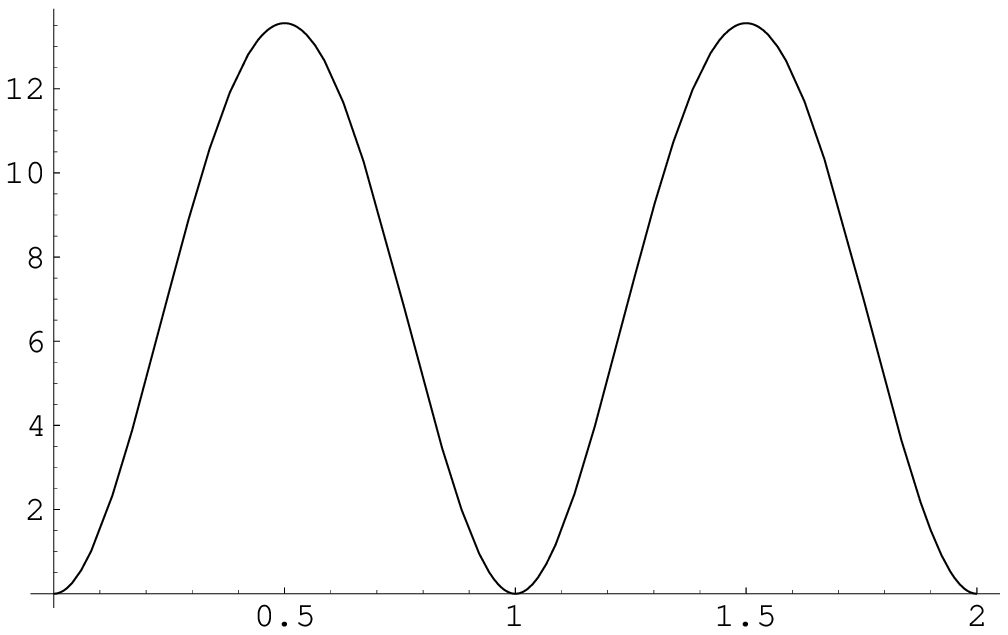}{The potential on two ${\tilde D}2$-branes
as a function of a non-Abelian tachyon vev.}

\EPSFIGURE{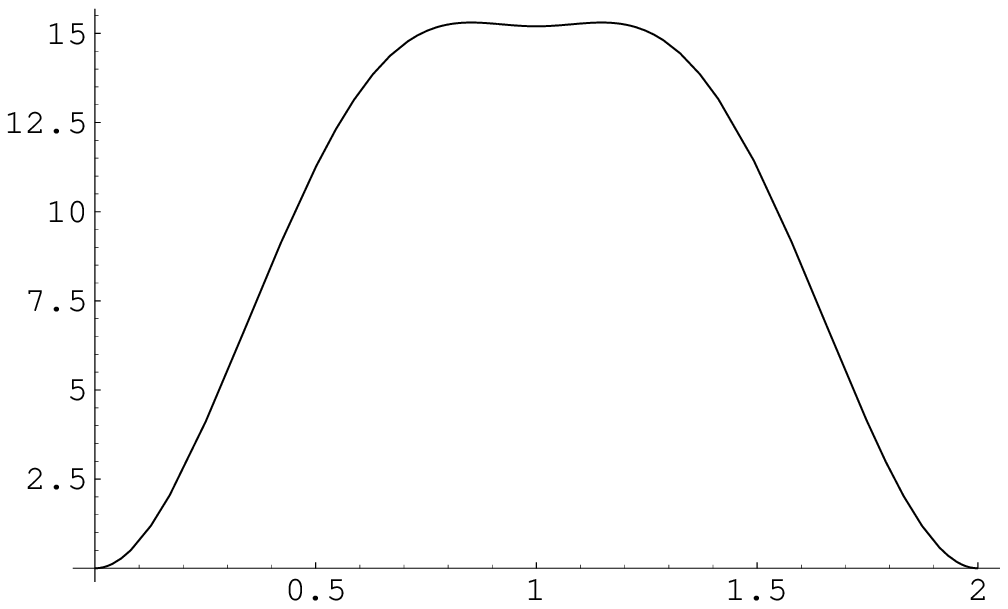}{The potential on two ${\tilde D}3$-branes
as a function of an Abelian tachyon vev.}

\EPSFIGURE{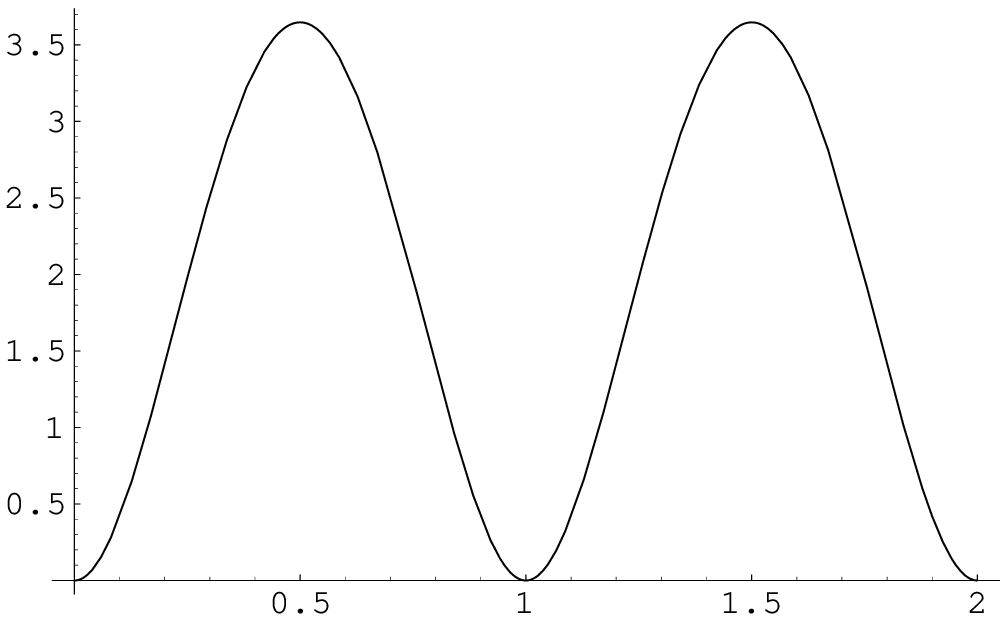}{The potential on two ${\tilde D}3$-branes
as a function of a non-Abelian tachyon vev.}

\EPSFIGURE{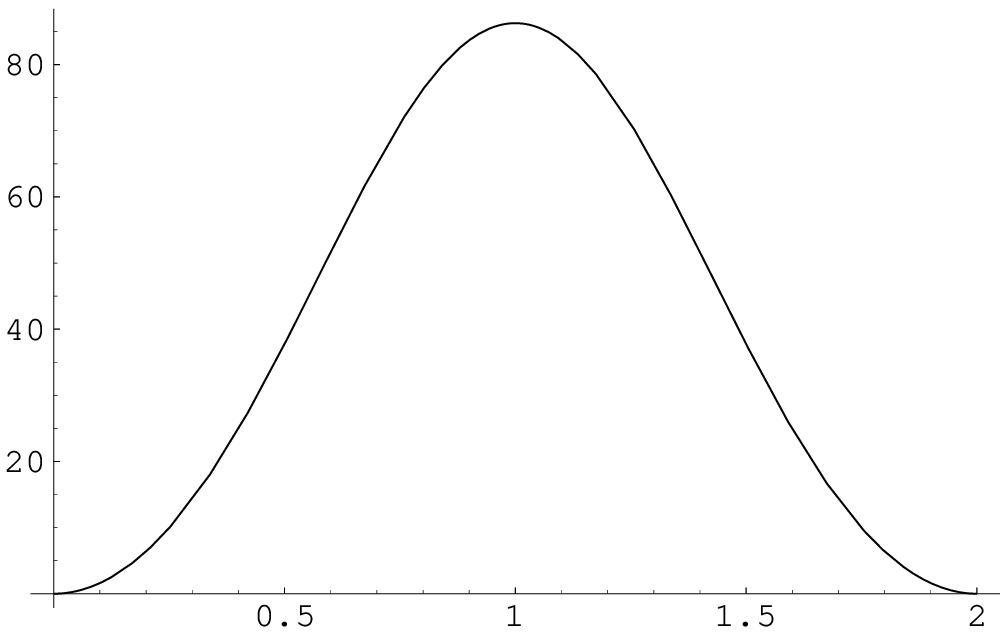}{The potential on two ${\tilde D}4$-branes
as a function of an Abelian tachyon vev.}

\EPSFIGURE{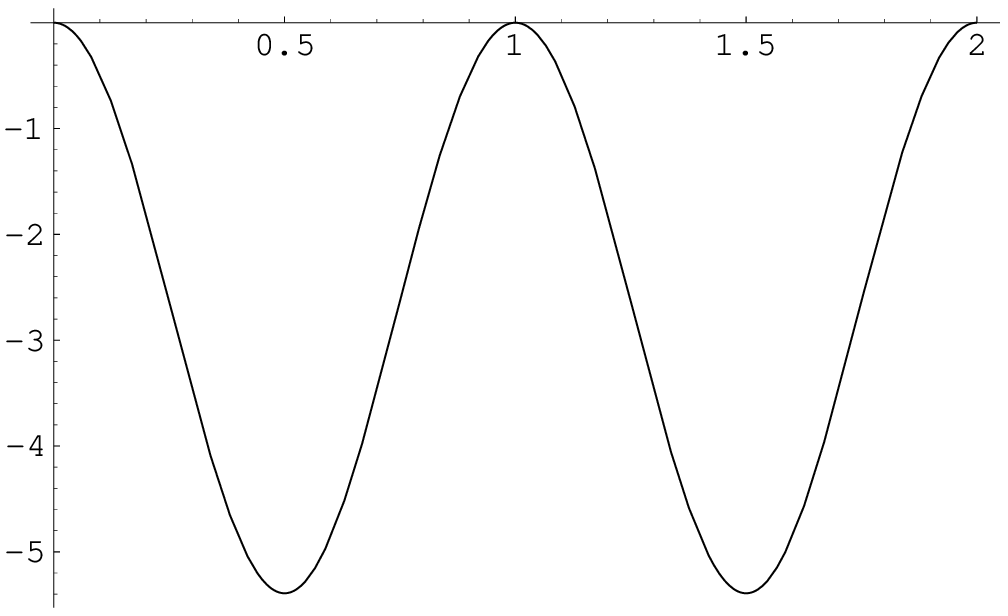}{The potential on two ${\tilde D}4$-branes
as a function of a non-Abelian tachyon vev.}

\end{document}